\documentclass[journal]{IEEEtran}
\usepackage{cite}
\usepackage{graphicx} 
\usepackage{array} 
\usepackage{bm} 
\usepackage{hyperref} 
\usepackage{color}
\usepackage{float}
\usepackage[dvipsnames]{xcolor}
\usepackage{physics}
\usepackage{soul}
\usepackage{ulem}
\usepackage{filecontents}

\graphicspath{{./}}
\usepackage{xcolor}
\usepackage{upgreek}

\usepackage{makecell}
\usepackage{amsmath,amssymb}

\usepackage{authblk}

\begin{document}
\bstctlcite{IEEEexample:BSTcontrol}

\title{Diamond Integrated Quantum Photonics: A Review}

\author{
 \IEEEauthorblockN{
 Prasoon K.\ Shandilya\IEEEauthorrefmark{1},
 Sigurd Fl\aa{}gan\IEEEauthorrefmark{1},
 Natalia C.\ Carvalho\IEEEauthorrefmark{1},
 Elham Zohari\IEEEauthorrefmark{2},
 Vinaya K.\ Kavatamane\IEEEauthorrefmark{1},\newline
 Joseph E.\ Losby\IEEEauthorrefmark{1},
 and Paul E.\ Barclay\IEEEauthorrefmark{1}
}\\
\IEEEauthorblockA{\IEEEauthorrefmark{1}Institute for Quantum Science and Technology, University of Calgary, Calgary, Alberta T2N 1N4, Canada}\\
\IEEEauthorblockA{\IEEEauthorrefmark{2}Department of Physics, University of Alberta, Edmonton, Alberta T6G 2E1, Canada}\\
\thanks{Corresponding author: pbarclay@ucalgary.ca}
}
\IEEEaftertitletext{\vspace{-1.3\baselineskip}}

\maketitle

\begin{abstract}
Integrated quantum photonics devices in diamond have tremendous potential for many quantum applications, including long-distance quantum communication, quantum information processing, and quantum sensing. These devices benefit from diamond's combination of exceptional thermal, optical, and mechanical properties. Its wide  electronic bandgap makes diamond an ideal host for a variety of optical active spin qubits that are key building blocks for quantum technologies. In landmark experiments, diamond spin qubits have enabled demonstrations of remote entanglement, memory-enhanced quantum communication, and multi-qubit spin registers with fault-tolerant quantum error correction, leading to the realization of multinode quantum networks. These advancements put diamond at the forefront of solid-state material platforms for quantum information processing. Recent developments in diamond nanofabrication techniques provide a promising route to further scaling of these landmark experiments towards real-life quantum technologies. In this paper, we focus on the recent progress in creating integrated diamond quantum photonic devices, with particular emphasis on spin-photon interfaces, cavity optomechanical devices, and spin-phonon transduction. Finally, we discuss prospects and remaining challenges for the use of diamond in scalable quantum technologies.
\end{abstract}

\section{Introduction}
\IEEEPARstart{Q}{uantum} technology has come a long way since the Stern–Gerlach experiment was conducted a century ago\cite{Gerlach1922}. Developments in quantum theory have provided the foundation for much of present-day technology, including, but not limited to, lasers, atomic clocks, and techniques in medical imaging, and have the potential to impact diverse fields spanning computing\cite{DiVincenzo1995, Ladd2010}, communication\cite{Gisin2007, Kimble2008}, and sensing\cite{giovannetti2004quantum, Giovanetti-PRL-2006-QuantumMetrology, Degen-QuantumSensing-RMP-2017}. We are now at an exciting moment where early-stage quantum computers have demonstrated quantum advantage\cite{Arute-Nature-2019-QuantumSupremacy,Zhong2020,Wu2021,Madsen2022}, quantum-secured communication has been demonstrated over intercontinental distances\cite{Liao-PRL-2018-SatelliteInrecontinentalQKD,Yin-SatelliteQKD-2020-Nature}, and quantum sensors have surpassed the sensitivity limits defined by the laws of classical physics\cite{kim2016approaching}.

A cornerstone for the development of real-world quantum applications is the resilient and scalable interconnection of different quantum systems. While several proof-of-principle experiments have been conducted with trapped atoms and ions\cite{Duan2010,Ritter2012,ReisererRempe-RMP-2015-CavityQED,Hosten2016,Stephenson2020,Egan2021,Langenfeld2021,Langenfeld2021PRLTeleportation,VanLeent2022}, scaling to a larger quantum system remains a challenge, in-part due to the complex laser trapping techniques required. Solid-state systems offer the possibility to adapt nanofabrication capabilities developed by the semiconductor industry to mass-produce quantum devices\cite{Koehl2015}. Superconducting quantum circuits\cite{Devoret2013}, gate-defined quantum dots\cite{ref:hanson2007sfe}, and semiconductor spin qubits\cite{burkard2021semiconductor} all take advantage of scalable nanofabrication to create quantum information processing devices. 
Integrated photonic devices could play a key role in these efforts by harnessing light's ability to transmit quantum information, and are becoming increasingly important as quantum technologies mature. Diamond has emerged as a promising material for photonic quantum systems build around artificial atoms formed by impurities in the diamond crystal\cite{Awschalom2018}. These atomic-scale impurities are optically active, allowing them to the controlled and connected through photonic channels, and to be interfaced with other photonic quantum technologies. 

In recent decades, the development of photonic quantum technologies\cite{OBrien2009} has expanded the quantum information ecosystem, particularly in the fields of quantum communication and quantum sensing. Quantum communication has many potential applications, including clock synchronization\cite{Komar2014} and encrypted communication\cite{Ekert-PRL-1991-QKD,Gisin-RMD-2002-QuantumCryptogrpahy}. Central to these applications is a quantum photonic network enabling the transmission and entanglement of quantum bits (qubits) over long distances\cite{Northup2014}. 

\begin{figure}[t!]
	\includegraphics[width=\linewidth]{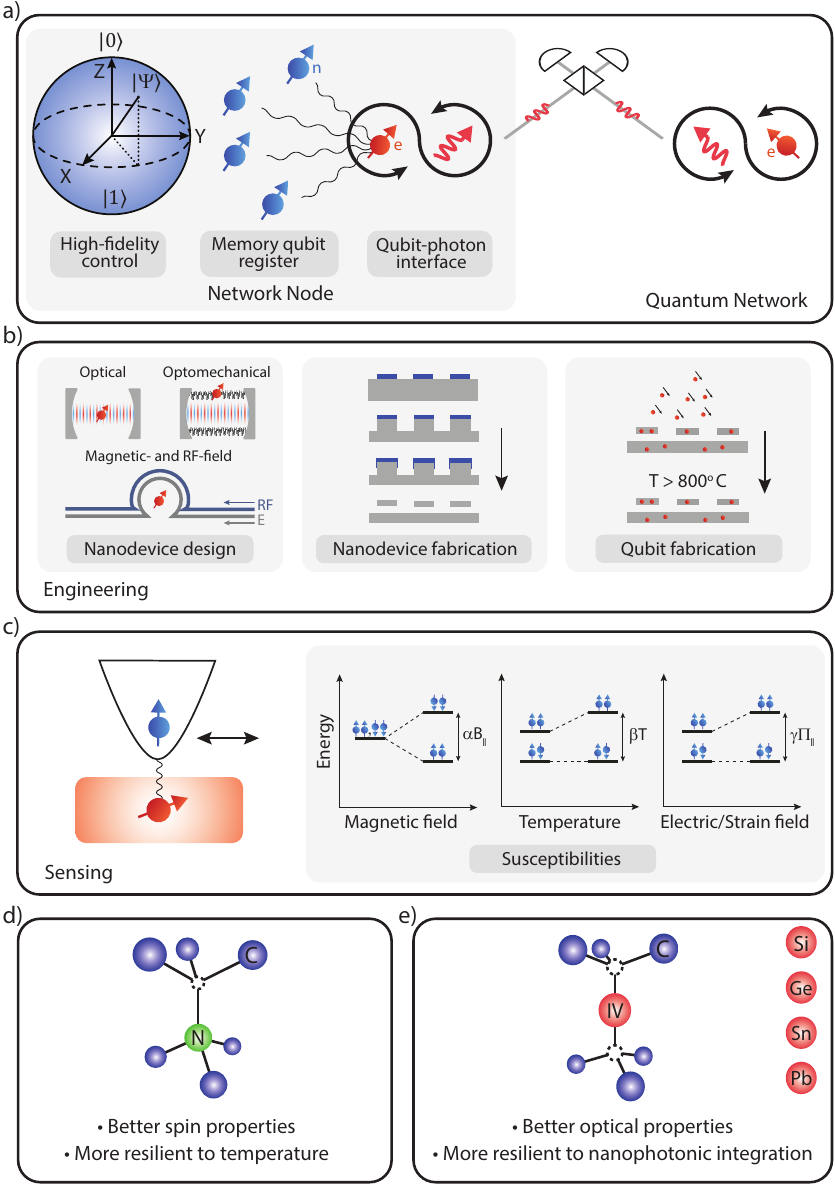}
		\caption{\textbf{Key motives for diamond integrated quantum photonics:} \textbf{(a)} The requirements of a quantum network node. A photonic platform in diamond is a promising candidate to meet all the requirements. \textbf{(b)} The solid-state structure of diamond allows for a variety of device designs utilizing photons, phonons and applied RF\,/\,magnetic fields to manipulate qubits via en-masse fabrication techniques. Color centers can be deterministically created through implantation and annealing. \textbf{(c)} A quantum sensor based on the single electron spin associated with the NV center. Sensing of the local spin environment or magnetic fields, temperature, and electric\,/\,stain field has been demonstrated using NV centers. Schematic structure of  \textbf{(d)} the NV and \textbf{(e)} the group-IV split-vacancy centers in diamond. The NV center has outstanding spin properties, with long coherence times at elevated temperatures. The inversion symmetry of the group-IV color centers leads to a larger Debye-Waller factor and a vanishing permanent electric dipole moment, making these color centers suitable for integration in nanophotonic devices. 
}
		\label{fig:summary}
\end{figure}

To realize a quantum network, remote quantum network nodes must be entangled via quantum links. As illustrated in Fig.\,\ref{fig:summary}\,(a), these network nodes are small-scale quantum processors that combine robust storage of quantum information with an interface to the quantum communication channels\cite{Wehner-Science-2018-QuantumInternetProtocols}. The nodes must enable high-fidelity operations on several entangled qubits and fault-tolerant multi-qubit protocols\cite{Abobeih2022}. Furthermore, the nodes need to be capable of storing quantum states utilizing `memory' qubits for a longer time than is required to generate asynchronous entanglement across numerous distant nodes. An additional requirement for network nodes is the realization of a fast and efficient interface between stationary qubits and `flying' photonic qubits. For long-distance communication, the flying qubits must have minimal losses in the photonic quantum channel, thus limiting the use to telecom photons that are compatible with pre-existing low-loss optical fiber networks\cite{Valivarthi2016,wandel2005attenuation,Northup2014}.

Optically active spin-qubits in diamond have successfully been utilized in several proof-of-principle experiments targeted towards the integration of the above mentioned components to realize a quantum network\cite{pfaff2014unconditional,Kalb-Science-2017-EntanglementDistillationNV,Evans2018,bhaskar2020experimental,Pompili2021Layer}. Advances in spin-photon entanglement\cite{Togan-SpinPhotonEntanglement-2010} paved the way for photon-mediated entanglement of remote qubits\cite{Bernien2013,Hensen-Nature-2015-LoopHoleFreeBell1p3km,Humphreys-Nature-2018-RemoteEntanglement}, culminating in the recent demonstration of a three-node quantum network\cite{pompili2021realization,Hermans2022}. While these demonstrations have been fruitful, it is worth noting that parallel advances have been made with other experimental platforms including trapped atoms and ions\cite{Ritter2012,Hofmann-Science-2012-EntanglementBetweenRb20mApart,daiss2021quantum,Moehring-Nature-2007-EntanglementAtDistance,Bluvstein2022,Graham2022}, superconducting resonators\cite{Blais2020,Krinner2022}, self-assembled quantum dots\cite{warburton2013single,stockill2017phase,Zhai2022}, and defect-based qubits in other wide-bandgap semiconductors and dielectrics\cite{Zhong2015Nature,Zhong2017Science,Dibos2018,Anderson2019,Kindem-Nature-2020-SingleShotReadoutREion,Lukin2020,Son2020,Bourassa2020,Wolfowicz2020,Babin2022,Lukin2022arxiv}.

Realizing an efficient interface between diamond qubits and photons at telecommunication wavelengths for fiber-based quantum channels remains a significant hurdle as diamond-based qubits do not intrinsically couple to telecom photons. The optically active spin qubits found in diamond typically emit photons in the visible range, thus requiring quantum frequency conversion to telecom wavelengths\cite{Dreau-PRAppl-2018-WavelengthConversionNV} to mitigate absorption loss in fiber optic links\cite{Northup2014}. The need for optical frequency conversion, however, can be circumvented by taking inspiration from the development of quantum transducers\cite{Regal-QuantumTransducer-JoP-2011} and utilizing the mechanical properties of diamond in combination with device engineering\cite{shandilya2021optomechanical}. Diamond is the stiffest material known to man with Young's modulus exceeding 1,200\,GPa\cite{Lee-SpinMechanicsReview-JoO-2017}, facilitating the fabrication of high-frequency mechanical resonators\cite{Mitchell-DiamondOptomechanicalResonator-Optica-2016,Burek-DiamondOptomechanicalCrystal-Optica-2016}. These mechanical resonators have the potential to be universal quantum transducers\cite{Lauk-QuantScienceTechnology-2020-PerspectiveQuantumTransduction,Chu2020}, capable of coupling a myriad of different quantum systems, including connecting superconducting qubit resonators and optical photons\cite{Fan2018SciAdv,Mirhosseini2020}. Photon-phonon and subsequent spin-phonon coupling in a mechanical resonator constitute a promising route to realize a spin-photon interface natively operating at telecom wavelengths\cite{stannigel2010optomechanical, shandilya2021optomechanical,Raniwala2022}. 
Furthermore, the high mechanical frequencies of these diamond resonators suppress the population of thermal phonons, facilitating preparing of the mechanical resonator to the quantum ground state.

Unleashing diamond's full potential as a material platform for quantum information processing is impaired by two preeminent challenges: deterministic creation of highly coherent qubits and scalable device fabrication. A key requirement for scaling beyond a few qubits is the precise fabrication of arrays of identical qubits separated by a few nanometers, a level of precision required to magnetically couple the electron spins to mediate multi-qubit operations\cite{Taminiau2014,rong2015experimental,chen2020optimisation}. Significant progress has been achieved in `top-down' qubit fabrication techniques, however it remains challenging to meet the necessary requirements\cite{bayn2015generation}. A promising approach to overcome this obstacle is `bottom-up' atomically-precise fabrication that has been successfully demonstrated in silicon\cite{fuechsle2012single}.  Alternatively, proposals for coupling qubits via a quantum optical bus require devices that combine deterministic creation of qubits with quantum control via externally applied fields\cite{ref:benjamin2009pfm}, such as optical, stress, magnetic, and radio and microwave frequency electric fields, see Fig.\,\ref{fig:summary}\,(b). The engineering of these control mechanisms requires robust device fabrication. Utmost care needs to be taken during device fabrication as the qubits are intrinsically sensitive to their surrounding environment and their properties can be degraded by their proximity to surfaces\cite{Chakravarthi2021}. However, by turning the problem around, their sensitivity to their environment (see Fig.\,\ref{fig:summary}\,(c)) can be exploited to realize quantum sensors\cite{Degen-QuantumSensing-RMP-2017} that offer advantages over their classical counterparts. Diamond-based sensors are being developed at a rapid pace and offers to fill the gap created by other candidates in terms of resolution, portability, and cost.

In this article, we focus on recent developments in the field of diamond integrated  photonics and its impact on the quantum technology ecosystem. The article is organized as follows: In Sec.\ II  we introduce the physics and the strengths and weaknesses of diamond colour centers. In Secs.\ III and IV, we provide an overview of the recently developed field of diamond cavity optomechanics and discuss the potential of photon-phonon coupling for quantum technology in diamond. In Secs.\  V, VI, VII, and VIII, we review the current state-of-the-art, discuss open questions, and provide a roadmap unravelling the promising future for diamond nanofabrication, quantum photonic devices, sensing and metrology, and qubit-photon interfaces.

\section{Color centers in diamond}
\label{sec:ColorCenters}

Diamond, an inherently transparent material, is host to over 200 optically active defect centers\cite{Zaitsev2001,Aharonovich2011}. These defects, known as color centers, can occur naturally or be created in the lab (see Sec.\,\ref{Sec:Defect_fabrication}). Some color centers  combine atom-like optical transitions with long-lived electron spins, resembling ions trapped in the diamond lattice\cite{Childress2014}. 
Diamond's large electronic bandgap ($5.47\,\textrm{eV}$) and resulting wide transparency window and large energy separation between colour center energy levels and the crystal's conduction and valance bands, combined with a low population of thermal phonons (Debye temperature$\,\Theta\sim 2000\,\textrm{K}$\cite{Tohei2006,Lee-SpinMechanicsReview-JoO-2017}) promotes coherent single-photon emission and long spin coherence times, making diamond an ideal host for solid-state qubits. Furthermore, the diamond lattice is largely composed of nuclear spin-0 $^{12}$C atoms (natural abundance $98.9\,\%$), thus suppressing background magnetic noise. Synthetic growth of diamond using isotopically purified starting material further reduces magnetic noise, leading to prolonged spin coherence times\cite{Balasubramanian2009}. The electron spin associated with color centers may be utilized as a qubit that, when interfaced with photons, can form part of a quantum network node. Furthermore, coupling to nearby long-lived $^{13}\textrm{C}$ nuclear spins\cite{Childress2006b} can be harnessed to build quantum memories\cite{Maurer2012} and multi-qubit spin registers\cite{Bradly-PRX-2019-NV10qubit1minuteCoherence}. Entanglement swapping between the electron spin and the long-lived nuclear quantum memory\cite{Dolde2013,Kalb-Science-2017-EntanglementDistillationNV}, can free up the communication qubit, enabling high-fidelity multi-qubit protocols\cite{Dutt2007}. For scalability, it is desirable to position the memory qubit in close spatial proximity to the communication qubit; a challenging task using $^{13}$C spins owing to the low natural abundance of only $1.1\,\%$\cite{fuchs2011quantum}. As an alternative, a quantum memory can be realized using the nuclear spins intrinsic to the color center\cite{fuchs2011quantum}. This latter approach carries the advantage that the memory qubit is automatically located adjacent to the communication qubit\cite{Toyli2010}.

Being robust single-photon sources\cite{Dolan2018,Knall2022}, the nitrogen-vacancy (NV)\cite{Doherty2013} and silicon-vacancy (SiV)\cite{Becker2020} centers have over the years gained traction as promising candidates for a variety of applications in photonic quantum technologies\cite{Johnson2017,Awschalom2018}.
NV centers have been used in various proof-of-principle experiments, including, but not limited to quantum sensing\cite{Rondin2014}, quantum teleportation\cite{pfaff2014unconditional}, quantum error correction\cite{waldherr2014quantum,Taminiau2014,Cramer2016}, demonstration of a multi-qubit quantum processor with coherence times approaching a second for the electron spin\cite{Abobeih-NatComm-2018-1sCoherenceTimeNV} and a minute for the nuclear spin\cite{Bradly-PRX-2019-NV10qubit1minuteCoherence}, fault-tolerant operations\cite{Abobeih2022}, and the realization of  multinode quantum networks\cite{pompili2021realization, Hermans2022}. On the other hand, SiV centers integrated with nanophotonic structures with exceptionally large cavity cooperativities have been used to demonstrate memory-enhanced quantum communication\cite{bhaskar2020experimental}. 
In addition, novel color centers based on the heavier group-IV atoms such as germanium (Ge), tin (Sn), and lead (Pb), show promising potential for quantum applications\cite{Bradac2019}. However, research on these defects is still in the early stages and will therefore not be discussed extensively in this review. 

\subsection{Defect Structure}
\label{Sec:CC_configuration}
The NV center, depicted schematically in Fig.\,\ref{fig:summary}\,(d), consists of a substitutional nitrogen atom and an adjacent lattice vacancy with the symmetry axis along the $\langle111\rangle$ crystal direction. The group-IV atoms, on the other hand, are too large to occupy a carbon site. Instead, the impurity atom takes up an interstitial position, flanked by a vacancy on either side\cite{Gali2013,Wahl2020}. This split-vacancy configuration, with the group-IV atom positioned at the inversion point of the diamond lattice, is shown schematically in Fig.\,\ref{fig:summary}\,(e). The centrosymmetric configuration results in a vanishing permanent electric dipole moment, rendering the group-IV color centers insensitive to linear Stark shifts\cite{Aghaeimeibodi2021,DeSantis2021}. The unpaired electrons associated with the neutral charge states, NV$^0$ and XV$^{0}$, where X refers to one of the group-IV atoms in Fig.\,\ref{fig:summary}\,(e), can efficiently capture an electron from the environment, forming the negative charge states, NV$^-$ and XV$^{-}$, respectively\cite{Manson2006,Hepp2014}. The negative charge state is the most studied for both classes of defects. However, it is worth mentioning that for the SiV center, careful doping and surface treatment have stabilized the neutral charge state SiV$^0$\cite{Rose2018,Zhang2022}. Similar results are yet to be reported for the heavier group-IV defects. Therefore, in the context of this review, we will be referring to the negative charge states, unless explicitly stated otherwise.

\begin{table*}[t!]
\renewcommand{\arraystretch}{2}
\caption{Overview of the Spin and Optical Properties of NV and SiV Color Centers in Diamond}
\label{table_NV_SiV}
\centering
\begin{tabular}{|c c c c c c c c|}
    \hline
    \makecell{Color \\ center} & \makecell{$\lambda_\text{zpl}$ \\ (nm)}  & \makecell{Radiative \\ lifetime (ns)} & \makecell{Quantum \\efficiency } & \makecell{Debye–Waller\\factor} & \makecell{GS $T_2$\\(Hahn-echo)} &\makecell{Spin-strain \\ coupling (per strain)} & \makecell{Orbital-strain \\ coupling (per strain)} \\
    \hline
    \hline
    NV$^-$ & 637 & $\simeq 12^1$  & $>$ 85$\,\%^2$ & 0.03$^3$ & \makecell{1.7\,ms$^4$ \\ (RT, 69\,mT)} & \makecell{21.5\,GHz$^5$ (GS; RT) \\ $\sim $290\,GHz$^5$ (ES; $<$ 10\,K)} &  \makecell{$\sim$ 1\,Hz$^5$ \\ (ES; $<$ 10\,K)} \\
    \hline
    \hline
    SiV$^-$ & 738$^6$ & \makecell{$\simeq 1.7-1.8$$^7$} &  1-10$\,\%^8$ & 0.7$^9$ & \makecell{0.3\,ms$^{10}$ \\ (0.1\,K, 160\,mT)} & N\,/A\,$^\dagger$ & \makecell{$\sim$ 1\,PHz$^{11}$\\ (GS; mK) } \\
    \hline
\end{tabular}
\\
$^1$\cite{Robledo2011NJP}, $^2$\cite{Radko2016}, $^3$\cite{Riedel2017}, $^4$\cite{ishikawa2012optical}, $^5$\cite{MacQuarrie2017}, $^6$\cite{Hepp2014}, $^7$\cite{Rogers2014,Sipahigil-Science-2016-SiVplatform}, $^8$\cite{Neu2012a,Sipahigil-Science-2016-SiVplatform}, $^9$\cite{Dietrich2014}, $^{10}$\cite{Sukachev2017}, $^{11}$\cite{Meesala2018}.\\ $^\dagger$Intrinsic spin-strain coupling is absent in SiV$^-$\cite{jahnke2015}. In certain regimes, spin-orbit interaction results in a spin-strain susceptibility of $\sim0.1\,\frac{\text{PHz}}{\text{strain}}$\cite{Meesala2018}. 
\end{table*}

\subsection{Spin and Optical Properties}
\label{Sec:spin_properties}

The NV center ground-state manifold is composed of an orbital singlet, spin-triplet state ($S=1$)\cite{Maze2011}, where, for zero magnetic field, the $m_{\textrm{s}}=0$ and the $m_{\textrm{s}}=\pm1$ spin sub-levels are split by $2.87\,\textrm{GHz}$ due to spin-spin interactions\cite{Doherty2013}. For the group-IV split-vacancies, the ground-state manifold is an orbital-doublet state, where spin-orbit interactions split the orbital branches by $\Delta_{\textrm{GS}}$\cite{Hepp2014,jahnke2015}. As will be discussed below, the value of $\Delta_{\textrm{GS}}$ directly determines the spin coherence times. A comparison between the key spin and optical properties for the NV and SiV center is summarized in Table\,\ref{table_NV_SiV}.

As a direct consequence of the difference in symmetry (Fig.\,\ref{fig:summary}\,(e)-(f)) and corresponding energy level structure, the NV and SiV centers exhibit different spin and optical properties, in turn determining in which areas they find applications. Due to the weak spin-orbit coupling\cite{Zhang2020} and the largely spin-free diamond lattice\cite{Balasubramanian2009}, the NV center electron spin is only weakly coupled to its environment, resulting in exceptionally long spin-lattice relaxation times ($T_1$) and spin-coherence ($T_2$) at room temperature\cite{Lee-SpinMechanicsReview-JoO-2017}. The long spin coherence time makes the NV center a workhorse in quantum sensing applications across a broad temperature range\cite{Hedrich2021,Scheidegger2022}. The SiV center, on the other hand, suffers from relatively short spin coherence times. In the case of the SiV center, the spin coherence time is limited by phonon-assisted population transfer between the two orbital branches (split by $\Delta_{\textrm{GS}}\sim48$\,GHz), resulting in coherence times limited by the orbital $T_1$ time\cite{jahnke2015,Becker2016a,Becker2018}. Cooling down to millikelvin temperatures suppresses the thermal phonon population, thereby increasing the spin coherence time\cite{Sukachev2017}. SiV coherence time-scales are typically in the range of 100\,ns at room temperature, extendable to a few ms at millikelvin temperatures, thus limiting the potential sensing applications of SiV centers.

The observation of two-photon quantum interference from spatially separated emitters\cite{Bernien2012,Sipahigil2012,Sipahigil2014}, a prerequisite for remote entanglement protocols\cite{Bernien2013,Hensen-Nature-2015-LoopHoleFreeBell1p3km}, requires coherent emission of indistinguishable photons. For NV centers, the photon flux is limited by the long radiative lifetime of $12\,\textrm{ns}$. Furthermore, emission into the zero-phonon line (ZPL) at $637\,\textrm{nm}$, quantified by the Debye-Waller (DW) factor, accounts for only $\sim3\,\%$ of the total emission\cite{Riedel2017}: the remaining $97\,\%$ is accompanied by phonons, resulting in a broad phonon-sideband (PSB) extending to $\sim800\,\textrm{nm}$. The SiV center, on the other hand, exhibits a comparatively short radiative lifetime of $1.8\,\textrm{ns}$\cite{Rogers2014,Sipahigil-Science-2016-SiVplatform}. Moreover, $\sim70\,\%$ of the emitted photons are directed along the ZPL at $738\,\textrm{nm}$\cite{Hepp2014,Dietrich2014}. However, the SiV center suffers from a low quantum efficiency (QE) of only $\sim0.1$\cite{Neu2012a,Sipahigil-Science-2016-SiVplatform} compared to a near-unity QE for the NV center\cite{Radko2016}.

In brief, the large difference in the DW factor between NV and SiV centers can be attributed to the symmetry of the defect\cite{Zhang2020}. In the case of the NV center, excitation from the ground-state (GS) to excited-state (ES), denoted by $^3A_2$ and $^3E$, respectively\cite{Maze2011}, shifts the charge distribution towards the N atom\cite{Gali2009,Gali2019}. This charge redistribution alters the equilibrium position of the nuclei, resulting in a reduced overlap between the ground- and excited vibronic states manifested by a large PSB\cite{Doherty2013,Zhang2020}. In comparison, the charge distribution between the ground ($^2E_{\textrm{g}}$) and excited state ($^2E_{\textrm{u}}$) of the SiV center remains similar; there is little change in the nuclear equilibrium coordinates. Consequently, optical emission from the SiV center occurs largely into the ZPL\cite{Zhang2020,Gali2013}. 

For both families of color centers, the presence of a magnetic field lifts the ground-state spin degeneracy: the spin sub-levels can be individually addressed using externally applied  microwave\cite{Jelezko2004,Pingault2017,Sukachev2017} or strain fields\cite{Barfuss-StrongMechanicalDrivinfNV-NatPhys-2015,Kolbl2019,Meesala2018}, enabling the formation of a qubit. Furthermore, both defects exhibit cycling optical transitions, enabling all-optical spin control\cite{Chu2015,Rogers2014b} and single-shot spin-readout\cite{Robledo2011,Sukachev2017} at cryogenic temperatures.

\section{Coherent Photon-Phonon Coupling in Diamond: Cavity Optomechanics}
\label{Sec:cavity_om}

Initial efforts to integrate diamond qubits with photonic devices have focused on photon-spin coupling, as discussed in previous reviews\cite{Schroder2016,Janitz2020} and addressed later in this review. Recently, efforts to harness spin-phonon coupling in diamond for realizing both on-chip and long-distance quantum coherent connections between diamond spin qubits have emerged\cite{Lee-SpinMechanicsReview-JoO-2017, Wang2020APL}. Central to these efforts is the field of cavity-optomechanics, which can provide coherent photon-phonon coupling in integrated photonic devices. Cavity optomechanics\cite{Kippenberg-2014-RMP-CavityOptomechanics, Safavi-Naeini2019}, has enabled quantum technologies in fields including sensing and precision measurement, studies of the quantum properties of mechanical objects, quantum memories, and quantum transduction. It has been investigated in a plethora of materials and designs, and has recently been shown to be a viable and promising route to control and read out spin qubits in diamond\cite{shandilya2021optomechanical}. This latter demonstration harnesses the convenient frequency matching between mechanical resonances of integrated devices and solid-state qubit spin transitions.

Cavity optomechanical systems use two distinct mechanisms to create and enhance the interaction between the optical field in a cavity and mechanical resonators: moving mechanical boundaries\cite{ref:povinelli2005ewb} and photoelasticity\cite{balram2014moving}. The mechanical displacement of vibrating cavity walls drives the moving boundary effect, whereas the photoelastic effect is a bulk response that manifests as strain-induced modulation of the refractive index. These light-matter interactions are quantified through the single-photon optomechanical coupling rate, $g_0$, calculated as the product of the optical frequency shift per unit displacement and the mechanical zero-point fluctuation amplitude of a mechanical resonator mode of interest. Experimental determination of $g_0$, in general, is realized by the parametric fitting of well-known optomechanical (OM) effects, such as the optical spring effect and the optomechanical \text{(anti-)} damping. A direct measure of $g_0$ is possible by comparing the OM transduction with an artificial phase-modulated tone, a method commonly used and described by Gorodetsky \textit{et al.}\cite{ref:gorodetksy2010dvo}. These measurements are corroborated by numerical simulations utilizing perturbation theory and the finite element method (FEM)\cite{primo2020}.

To be used in quantum applications, optomechanical cavities must be designed to achieve high $g_0$ while also overcoming optomechanical losses. This requirement is characterized by the optomechanical cooperativity, given by:
\begin{equation}
C_\text{om} =  \frac{4 \overline{n}_{\text{cav}} g_0^2}{\kappa\gamma_{\text{m}}}\,.
\end{equation}
Here, $\kappa$ and $\gamma_\text{m}$ are the energy decay rates of the cavity's optical and mechanical modes of interest, respectively, and $\overline{n}_{\text{cav}}$ is the average intracavity photon number. $C_\text{om}$ can be interpreted as the probability of coherent OM interaction\cite{Borregaard2019}. The condition $C_\text{om}>1$ has been realized in a variety of cavity OM devices, including those fabricated from diamond\cite{Burek-DiamondOptomechanicalCrystal-Optica-2016, Mitchell-DiamondOptomechanicalResonator-Optica-2016}. 
Moreover, quantum state transfer between light and mechanics that can generate entanglement between photons and phonons demands devices with $C_\text{om}>n_\text{th}+1$, where $n_\text{th}$ is the average thermal phonon number of the mechanical resonator due to its thermal bath\cite{Cohen-PhononCounting-Nature-2015, Wallucks-QuantumMemoryTelecomBand-NatPhys-2020}. Lowering $n_\text{th}$ either requires cooling  resonators to cryogenic temperatures, using high-frequency resonators, or a combination of  both. Another condition that is key to many applications is reaching the sideband-resolved regime, that is, when the resonator's mechanical mode frequency exceeds $\kappa/2$, facilitating delayed optical back-action that can be harnessed for optical amplification or cooling of mechanical resonance. 

The above discussion only accounts for the real part of the optomechanical coupling. In cases where the change in a cavity's optical losses via mechanical displacement is significant, $g_0$ will be defined as a complex quantity with dispersive and dissipative contributions\cite{primo2020}. Furthermore, when the device is operated under high optical power, dynamic and static photothermal and thermoelastic effects may be relevant, which can significantly modify the physics of the problem\cite{primo2021, pinard2008quantum, ref:barclay2005nrs}. In this review, we focus on the dispersive OM interaction; the above-mentioned effects are neglected unless explicitly stated. 

\subsection{Relevant material properties}
\label{Sec:Material_Properties}
A fortuitous combination of exceptional intrinsic material properties is responsible for making diamond an ideal material for cavity optomechanics. 
Advances in the synthetic growth of single-crystal diamond (SCD) combined with the development of fabrication techniques to realize suspended structures from high-quality SCD chips have accelerated the recent emergence of diamond nanophotonic and nanomechanical devices for quantum technologies. Diamond's key material properties relevant to quantum photonics are outlined below.

\subsubsection*{Optics} 
The strong covalent bond of diamond hinders electrical conductivity, resulting in a wide bandgap of 5.47 \,eV\,(225\,nm). This large bandgap creates a broad transparency window ranging from ultraviolet to radio frequencies, except for a weak absorption window appearing between 2.6 and 6.2\,$\mu$m due to multi-phonon coupling\cite{Mildren2013ch1}. Outstanding transparency, when combined with a high refractive index of $\sim$ 2.4 and low dispersion, makes diamond a robust and versatile medium to confine photons. Additionally, the large bandgap suppresses multi-photon absorption, providing the ability to handle large optical power inside a cavity with minimal nonlinear absorption and resulting thermal effects, unlike other popular semiconductors such as silicon and gallium arsenide\cite{almeida2017nonlinear}.

While thermal nonlinear effects are mostly undesirable, optical nonlinearities lead to interesting physics. Diamond's lowest-order non-zero nonlinear susceptibility, $\chi^3$, allows the study of  phenomena such as Raman scattering\cite{eesley1978coherent, mildren2008cvd, kasperczyk2015stokes}, the Kerr effect\cite{okawachi2017competition, motojima2019giant}, and four-wave mixing\cite{lu2014generation}. Amid many prospects, the on-chip demonstration of  diamond-based microcombs\cite{lux2014multi}, Raman lasers\cite{Latawiec2015chip,Latawiec2018integrated}, and single-photon frequency conversion via a combination of parametric processes with color centers are among the most promising\cite{hausmann2014diamond}.

Special attention must be paid to surface termination, which can introduce additional losses when device dimensions reach the nanoscale. A high concentration of impurities, such as color centers, may have an impact on both optical and mechanical loss rates. Even though diamond is an optically isotropic material due to its cubic lattice symmetry, the strain induced by the presence of impurities can cause birefringence.

\subsubsection*{Mechanics} 

Diamond has exceptional mechanical properties. It has a high Young's modulus of 1220 GPa, four times larger than silicon, making it a stiff material with a large speed of sound ($\sim$18,000\,m\,/\,s). The large stiffness facilitates the fabrication of high-frequency resonators by reducing reliance on small geometry, ameliorating access to mechanical resonances resonant with a wide range of spin qubit transition frequencies. 
Equally important, diamond is a material with low intrinsic mechanical dissipation. In general, the predominant geometry-independent dissipative channels for nanomechanical semiconductor resonators near room temperature are thermoelastic damping and phonon-phonon scattering\cite{Bachtold2022}. Both these effects are temperature-dependent, and therefore, their contribution is relatively small in diamond due to its high thermal conductivity (2,200\,W\,/\,mK, several times higher than copper) and density (3,500\,kg\,/m$^3$), alongside its low thermal expansivity. At cryogenic temperatures, scattering by defects such as dangling bonds, intrinsic or extrinsic dopants, and lattice imperfections can dominate phonon losses\cite{Tao2014}. Such interactions are commonly modelled as strain waves coupling to two-level systems and might be the ultimate limit for achieving ultra high-Q nanomechanical resonators\cite{adiga2009mechanical}.

Besides the lattice anisotropy introduced by defects, even undoped SCD is an imperfect isotropic elastic material. Although the anisotropy level is often disregarded without penalty, fine tailoring of diamond's mechanical properties is sometimes necessary. Huang \textit{et al.}\cite{huang2018anisotropy} have demonstrated that Young's modulus varies depending on the crystal orientation. Likewise, other properties such as photo-elasticity are also influenced by the crystallographic direction, as demonstrated in gallium arsenide\cite{balram2014moving}. These variations can be accounted for by a thorough description of the diamond elasticity tensor\cite{lang2009strain,Barfuss2019}.

\subsection{Experimental realizations of diamond optomechanical devices}

\begin{figure*}[t!]
	\includegraphics[width=\textwidth]{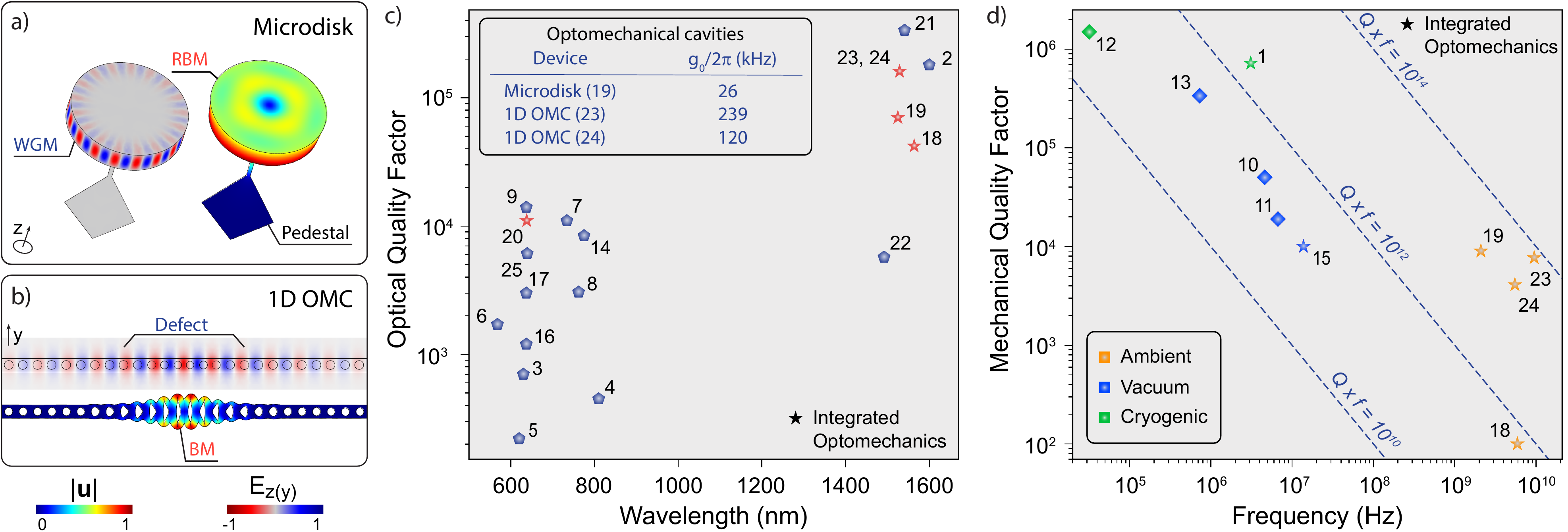}
		\caption{\textbf{Diamond optical and mechanical resonators:} Simulation of \textbf{(a)} microdisk cavity and \textbf{(b)} 1D optomechanical crystal (OMC). WGM: whispering gallery mode; (R)BM: (radial) breathing mode; $|\mathbf{u}|$ and $E_{z (y)}$ stand for normalized mechanical displacement and electric field at the z (y) direction, respectively. Reported quality factors: \textbf{(c)} optical and \textbf{(d)} mechanical diamond resonators (experimental data). Nanobeams: 1\cite{Khanaliloo2015PRX} and 11\cite{ref:burek2013nrs}; 1D Photonic crystals (PhC): 2\cite{Burek2014}, 3\cite{Riedrich-Moller2012}, 5\cite{bayn2011triangular}, 6\cite{li2015one}, 7\cite{Burek2017}, 8\cite{ref:bayn2014ftn}, 9\cite{Mouradian2017}, 14\cite{regan2021nanofabrication}, 18\cite{cady2019diamond}, 23\cite{Burek-DiamondOptomechanicalCrystal-Optica-2016} and 24\cite{Burek-DiamondOptomechanicalCrystal-Optica-2016}; 2D PhCs: 4\cite{Riedrich-Moller2012}, 15\cite{rath2014diamond}, 16\cite{Riedrich-Moller2015}, 17\cite{Faraon2012} and 25\cite{wan2018two}; Cantilevers: 10\cite{ref:burek2013nrs}, 12\cite{Tao2014} and 13\cite{ovartchaiyapong2012high}; Microdisks: 19\cite{Mitchell-DiamondOptomechanicalResonator-Optica-2016}, 20\cite{Mitchell-DiamondOptomechanicalResonator-Optica-2016}, 21\cite{Mitchell-2019-APLphotonics-DiamondMicrodisks} and 22\cite{graziosi2018single}.}
		\label{omfig}
\end{figure*}

Efforts to utilize optical fields to manipulate and probe micro-and nano-scale mechanical objects have led to numerous types of optomechanical devices\cite{Safavi-Naeini2019}. Amongst these, microdisk resonators are one of the simplest optical cavity designs and play an important role in many photonics and optomechanics experiments. Microdisks support optical \textit{whispering gallery modes} (WGM), in which the electromagnetic field is confined by total internal reflection at the circular boundary (Fig.\,\ref{omfig}\,(a)). 
The mode volume of WGMs decreases with decreasing microdisk radius, while optical radiation loss increases.
An optical quality factor of $Q_\text{o} \sim 10^{5}$ at telecommunication wavelengths was achieved in 5$\,\mu\text{m}$ diameter SCD microdisks by Mitchell \textit{et al.}\cite{Mitchell-2019-APLphotonics-DiamondMicrodisks}, limited by fabrication related surface roughness. At visible wavelengths resonant with NV optical transitions, $Q_\text{o} > 10^4$ have been observed in SCD microdisks\cite{Mitchell-DiamondOptomechanicalResonator-Optica-2016}.

Optomechanical microdisks are supported by small pedestals that minimize anchor losses, as shown in Fig.\,\ref{omfig}\,(a). Optical WGMs couple most effectively to mechanical radial breathing modes (RBM): resonances with mechanical frequency $\Omega_\text{m}/2\pi\sim2$\,GHz, mechanical quality factor $Q_\text{m} \sim 10,000$, and $g_0/2 \pi \simeq 26$\,kHz have been reported for a diamond microdisk\cite{Mitchell-DiamondOptomechanicalResonator-Optica-2016}. This device demonstrated $C_\text{om}\sim 3$ at ambient conditions, enabling a number of coherent optomechanics experiments, including reversible photon-phonon conversion via optomechanically-induced transparency\cite{Lake-OMIT-diamond-microdisks-ACSPhotonics-2018, lake2021processing}, phonon-mediated wavelength conversion\cite{Mitchell-Optica-2019-OptomechanicsWavelengthConversion}, and optomechanical memory\cite{lake2021processing}. These demonstrations took advantage of the multimode nature of the diamond microdisk; a topic that will be discussed in Sec.\,\ref{Sec:mm_OM}. 

The advantages of microdisk resonators go beyond their ease of fabrication and effective OM interaction. Usually, resonators with small dimensions are targeted due to their low effective modal mass, leading to a higher $\Omega_\text{m}$ and larger $g_0$.  Tuning of the device geometry modifies the optical free spectral range (FSR) and can enable mechanical frequencies ranging from kHz to GHz -- such design flexibility is favorable for coupling to spin qubits\cite{Wang2020APL}. On the downside, mechanical quality factors are highly dependent on the disk-to-substrate interconnection. While M.\ Mitchell \textit{et al.}\cite{Mitchell-DiamondOptomechanicalResonator-Optica-2016} demonstrated the feasibility of fabricating a thin pedestal in diamond microdisks, the patterning of a nanostructured phononic shield on the top of the disk could provide a solution to further isolate clamping and reduce anchor losses\cite{carvalho2021high, santos2017hybrid}.

Optomechanical crystals (OMC) are another device design used to simultaneously confine optical and mechanical modes\cite{Eichenfield--OptomecanicalCrystal-Nature-2009}. Their sophisticated engineering allows for small optical mode volumes and high OM coupling rates compared to other geometries. The basic principle involves introducing a geometric defect in an artificially patterned lattice (Fig.\,\ref{omfig}\,(b)) that generates periodicity in the dielectric constant and elastic properties of a material. This in turn modifies the dispersion of photonic and phononic modes of the structure. These defects are generally implemented by alternating air and material regions, for example by introducing air holes in a suspended film. The resulting photonic and phononic band structures have regions with vanishing density of states. Introducing defects that break the lattice's translational symmetry allows the formation of localized optical and mechanical resonances at frequencies within the bandgaps of the respective optical and photonic band structures. Additionally, these modes can simultaneously have low radiation and clamping losses, respectively, by ensuring that the defect minimizes the coupling to the lossy modes of the underlying periodic structure\cite{joannopoulos2011photonic}. 

Diamond photonic crystal cavities (PCC) and nanomechanical resonators were first explored independently. Photonic crystal optical cavities using triangular nanobeam PCCs were demonstrated with a $Q_{\text{o}}\,\sim\,1000$\cite{bayn2011triangular}, soon followed by a $Q_\text{o} > 10^5$ due to advances in fabrication techniques\cite{Burek2014}. Rectangular cross-section photonic crystal cavities\cite{li2015one} better suited for integration into a larger photonic integrated circuit have been demonstrated using quasi-isotropic undercutting of SCD with $Q_\text{o} \sim 14,000$\cite{Mouradian2017}. These rectangular cross-section cavities minimize the mixing of the TE and TM optical modes which can be responsible for optical loss in triangular cross-sections\cite{bayn2011triangular}.

Photonic crystal cavities fabricated using heteroepitaxy of diamond  on silicon\cite{Riedrich-Moller2012} or diamond membrane transfer\cite{regan2021nanofabrication} have $Q_\text{o} \sim 10^4$. Two-dimensional PCCs, have been realized from diamond films bonded or deposited over a different material substrate. They can currently reach optical $Q$-factors of  a few thousand\cite{Riedrich-Moller2012, rath2014diamond, Riedrich-Moller2015, Faraon2012}. Recently, quasi-isotropic etching\cite{Khanaliloo2015} was used to realize 2D photonic crystals with $Q_\text{o} > 6\times 10^3$\cite{wan2018two}. Several of the aforementioned devices have been realized with cavities in the visible wavelength range, as shown in Fig.\,\ref{omfig}\,(c). 

The realization of SCD nanomechanical resonators has made steady progress in recent years, as illustrated in Fig.\ \ref{omfig}\,(d). Cantilevers with $\mu\text{m}$ dimensions suspended over a SiO$_2$ substrate have been demonstrated with a $Q_{\text{m}} \sim 3 \times 10^5$ in vacuum\cite{ovartchaiyapong2012high}, while Tao \textit{et al.}\cite{Tao2014} reported that similar structures made from SCD film in a `quartz sandwich' can achieve  $Q_{\text{m}} \sim 10^6$. These high-$Q_\text{m}$ cantilevers have also been used to investigate the relevance of surface termination and defect concentration. In Ref.\cite{ref:burek2013nrs} both singly- and doubly-clamped nanobeams showed $Q_{\textrm{m}}>10^4$ in angle-etched SCD.

Khanaliloo \textit{et al.}\cite{Khanaliloo2015PRX} demonstrated one-dimensional suspended waveguides in SCD supporting flexural mechanical modes with $Q_{\text{m}} > 7 \times 10^5$. Polycrystalline diamond nanobeams and H-resonators with $Q_{\text{m}} \sim 10^4$ have been connected to nanophotonic circuits\cite{ref:rath2013doc,rath2014diamond}. Cady \textit{et al.}\cite{cady2019diamond} have fabricated diamond OMCs using a SCD-on-insulator platform with integrated telecom waveguides with $Q_{\text{o}}>10^4$ and GHz mechanical modes with $Q_\text{m}\sim 100$. Burek \textit{et al.}\cite{Burek-DiamondOptomechanicalCrystal-Optica-2016} fabricated and measured a 1D OMC with a  $Q_\text{o} > 10^5$, $Q_{\text{m}} \times f_{\text{m}} \sim 10^{14}$ and $g_0/2\pi \sim 240$ kHz. These realizations demonstrate that diamond cavity optomechanics has enormous potential. However, the fabrication of smooth sidewalls and the precise etching of small periodic features in diamond OMCs are currently major obstacles to further experimental progress, as will be discussed in Sec.\,\ref{Sec:Dfab}.

\section{Multimode Optomechanics}
\label{Sec:mm_OM}

\begin{figure}[t!]
	\includegraphics[width=\linewidth]{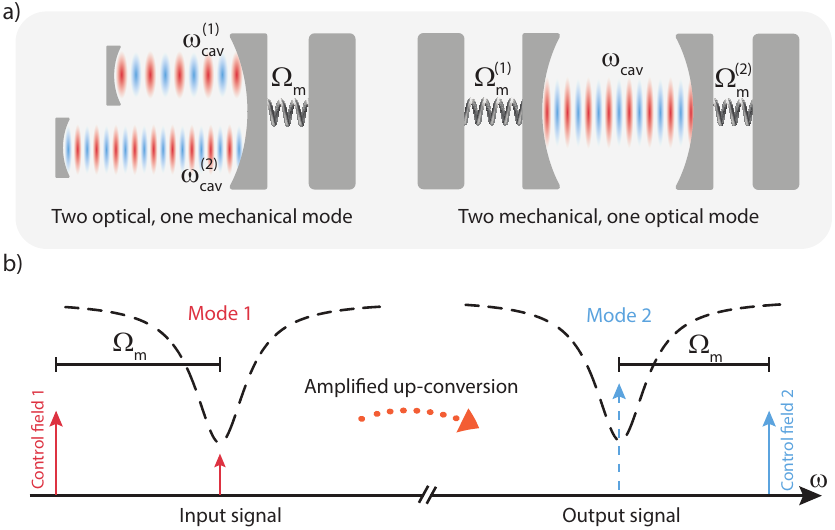}
		\caption{\textbf{Multimode cavity optomechanical systems and applications: }\textbf{(a)} Schematic illustration of two different multimode optomechanical systems. \textbf{(b)} Amplified wavelength conversion using an optomechanical device. Two optical modes at different frequencies are illuminated by a strong red-detuned and blue-detuned control beam, respectively. 
		A signal injected at the resonance of mode 1 is converted into a signal emerging from mode 2, as demonstrated using diamond microdisk in Ref.\,\cite{Mitchell-Optica-2019-OptomechanicsWavelengthConversion}.}
		\label{fig:mm_OM}
\end{figure}

The majority of  cavity optomechanics studies utilize systems with a single optical mode coupled to one mechanical mode. As discussed, an asset of diamond cavities is their ability to support optical modes across a wide transparency window, thus facilitating the study of multimode optomechanics, where multiple optical modes are coupled to the same mechanical mode (see Fig.\,\ref{fig:mm_OM}\,(a, left)). Similarily, mechanical resonators typically support a spectrum of modes that can couple to a single optical mode (see Fig.\,\ref{fig:mm_OM}\,(a, right)). Constraining the physics to the `one-to-one' model is often justified and dictated by the experimental design. For example, cooling or heating in the resolved sideband regime singles out a particular mechanical mode via the choice of laser detuning\cite{ref:teufel2011scm}. 

Setups with multimode coupling have been realized in a wide range of materials\cite{ref:thompson2008sdc} and have a number of additional valuable features\cite{woolley2013two, kuzyk2017controlling, ockeloen2016quantum, pontin2016dynamical, nielsen2017multimode, dong2014optomechanically}. For example, bulk acoustic modes in a macroscopic scale quartz cavity have been used to demonstrate optomechanically induced transparency and absorption\cite{kharel2019high}. Multiple mechanical modes coupled to an optical mode have demonstrated a cascaded optical transparency scheme by leveraging the parametric phonon–phonon coupling with an aluminium nitride microwheel resonator\cite{fan2015cascaded}, coherent optomechanical state swapping between two spatially and frequency separated resonators with silicon nitride trampoline resonators\cite{weaver2017coherent}, and multimode phonon lasing with a 1D silicon optomechanical crystal cavity\cite{mercade2021floquet}. Furthermore, exploring multimode systems in synchronized OM oscillator arrays constitutes an interesting approach for sensing and metrology\cite{zhang2015synchronization}.

Multimode cavity optomechanical devices also offer great potential to connect disparate quantum technologies in a network, via phonon-mediated coherent transduction of quantum information from visible or microwave photons to telecommunication wavelengths. As such, the field of optical-to-microwave transduction\cite{Forsh-MwtoOpticsTransduvtionOptomechanics-NatPhys-2020} is currently a highly active area of research\cite{Lambert2020coherent}. Optomechanical systems with $C_\text{om} > 1$ are capable of frequency conversion between multi-wavelength cavities\cite{Hill-NatComm-2012-OptomechanicalWavelengthConversion, liu2013electromagnetically, Mitchell-Optica-2019-OptomechanicsWavelengthConversion}, eliminating the need for material-dependent nonlinear optical processes. Cavity optomechanical wavelength conversion coherently couples two optical cavity modes via their independent optomechanical coupling to a common mechanical resonance\cite{Hill-NatComm-2012-OptomechanicalWavelengthConversion, Mitchell-Optica-2019-OptomechanicsWavelengthConversion}.

In diamond, optomechanical cavities have been used to demonstrate wavelength conversion between frequencies separated by over 4\,THz with an internal efficiency of 45\,$\%$\cite{Mitchell-Optica-2019-OptomechanicsWavelengthConversion}. For wavelength up- and down-conversion, both strong optical control fields are typically red-detuned by the mechanical frequency from cavity modes resonant with each wavelength of interest. However, blue detuning of one of the control fields can be harnessed for optomechanically amplified wavelength conversion, as illustrated in Fig.\,\ref{fig:mm_OM}\,(b). Dynamics in both of these wavelength conversion schemes can be modelled by a photon-phonon beamsplitter or squeezing Hamiltonians\cite{Kippenberg-2014-RMP-CavityOptomechanics}. These systems can, in principle,  be utilized to convert the zero phonon line  emissions from NV and SiV centers in diamond to telecom photons for application in quantum networks, provided the cavity supports high-$Q_\text{o}$ optical modes at both the photon emission wavelength and the telecom wavelength of interest, and that $C_\text{om} \ge 1$ can be achieved for each mode. However, these advancements are currently limited due to modest $Q_\text{o}$ at visible wavelengths. Advances in fabrication techniques and better cavity design can overcome this problem, see Sec. \ref{Sec:Dfab}. Coupling higher intensity control fields to the cavity at the visible wavelength in order to achieve larger $\overline{n}_\text{cav}$ will also increase $C_\text{om}$.

Other multimode optomechanical phenomena with promising quantum applications have been demonstrated in diamond devices enabled by double-optomechanically induced transparency (DOMIT). Examples include all-optical two-color switching\cite{lake2020two} and optomechanicaly tunable pulse storage\cite{lake2021processing}. Recently, optical modes were used to drive a broad mechanical mode spectrum of diamond microdisks. In this work,  higher-order mechanical modes with frequencies over 10\,GHz were observed. Devices with one optical and multiple mechanical modes could lead to many interesting studies, as already demonstrated in the microwave domain\cite{massel2012multimode}, but have not yet been explored in SCD optomechanical systems. Hybrid SCD cavity optomechanical devices integrated with superconducting quantum devices could accelerate the realization of transducers for connecting superconducting quantum computers to quantum networks\cite{Lauk-QuantScienceTechnology-2020-PerspectiveQuantumTransduction}.

\section{Diamond Nanofabrication}
\label{Sec:Dfab}
Advances over the last decade in the fabrication of nanostructures from single-crystal diamond (SCD) have accelerated the development of components necessary for fully integrated quantum photonics platforms.  These components, engineered to enhance light-matter interactions, play a crucial role in the fields of cavity optomechanics (see Sec.\,\ref{Sec:cavity_om} and \ref{Sec:mm_OM}), nanophotonics (see Sec.\,\ref{sec:nanophotonicdevices}), and qubit-photon interfaces (see Sec.\,\ref{Sec:interface}). They include, but are not limited to, single-photon sources based on color-centers\cite{Aharonovich2011}, wave-guiding structures\cite{Hadden2018}, optical\cite{Faraon2011,Faraon2012} and mechanical resonators\cite{Mitchell-DiamondOptomechanicalResonator-Optica-2016,cady2019diamond}, and optical fiber-based\,/\,free-space couplers\cite{Burek2017,Dory2019NatCom,Chakravarthi2020Optica}. In this section, the challenges of SCD nanostructuring are briefly introduced, followed by discussions of top-down fabrication methods for bulk diamond, as well as methods for creating color centers in these structures.

\subsection{Single-Crystal Diamond Fabrication Challenges}

In addition to diamond's ability to host color centres, its allure for nanophotonic and nanomechanical applications derives from the advantageous material properties of diamond introduced above (see Sec.\ref{Sec:Material_Properties}).
The unit cell of SCD consists of two diagonally inter-penetrating face-centered cubic lattices where each carbon atom is purely covalent-bonded to its four nearest-neighbours with tetrahedral symmetry. This atomic structure grants diamond its unique mechanical, thermal, and optical properties, including the wide transparency window and chemical inertness\cite{Mildren2013ch1}.

Although these properties make SCD an excellent material for integrated photonic devices, they also provide significant challenges in nanostructuring. Traditional top-down nanofabrication approaches rely on lithographic patterning of masking material, and subsequent etching steps to define photonic and mechanical components.  The high chemical inertness of diamond limits etching chemistries to processes involving oxygen plasma, while the strong bonding nature of the carbon atomic lattice makes physical etching, such as focused ion beam milling, difficult.  Currently, high-quality SCD in thin-film form, where a refractive index contrast confines light within the film thickness, is not commercially available, although recent progress has been made in hetero-epitaxially grown films on various substrates\cite{Kim2021DiamondGrowth}.  Though polycrystalline thin-films are readily available, they might not be suitable for hosting quantum emitters due to the presence of dopants and grain boundaries\cite{Mildren2013ch1,rath2015diamond}. Several paths have been taken in order to overcome these challenges, both in etch mask development and through the use of different SCD precursor materials, ranging from thinned membranes to bulk substrates.

\subsection{Creating suspended structures in bulk diamond}
High-quality, commercially available SCD in bulk form, grown on diamond templates via chemical vapour deposition (CVD), is a popular precursor material for diamond nanostructuring.  Both optical and electronic grade forms of CVD diamond, with the latter having less than 10 parts per billion in residual nitrogen density, are available down to millimeter-scale substrate form.  The substrate surfaces are mechanically polished (a specialized procedure that is often outsourced) to sub-nanometer RMS surface roughness.  Initial cleaning of the substrates typically uses a boiling piranha solution (3:1 ratio of sulfuric acid and hydrogen peroxide) for approximately 15 minutes, followed by a rinse in water and drying with a nitrogen gun.  These steps can be supplemented with a combination of a hydrofluoric acid dip for 5 minutes and sonication in acetone and or methanol\cite{atikian2014superconducting}.  Optical microscopy inspection of the substrate post-cleaning should yield a clean surface, absent of graphitic or other residues, often evident as dark spots.  If not, the cleaning process should be repeated.  In cases where the substrates have been treated with high-temperature annealing or ion irradiation, the induced graphitic or pyrolytic carbon residue can be etched in a tri-acid cleaning solution (a boiling mixture of nitric, sulfuric, and perchloric acids), though care should be taken with the associated risks in this potentially volatile process.

\begin{figure*}[t]
	\includegraphics[width=\textwidth]{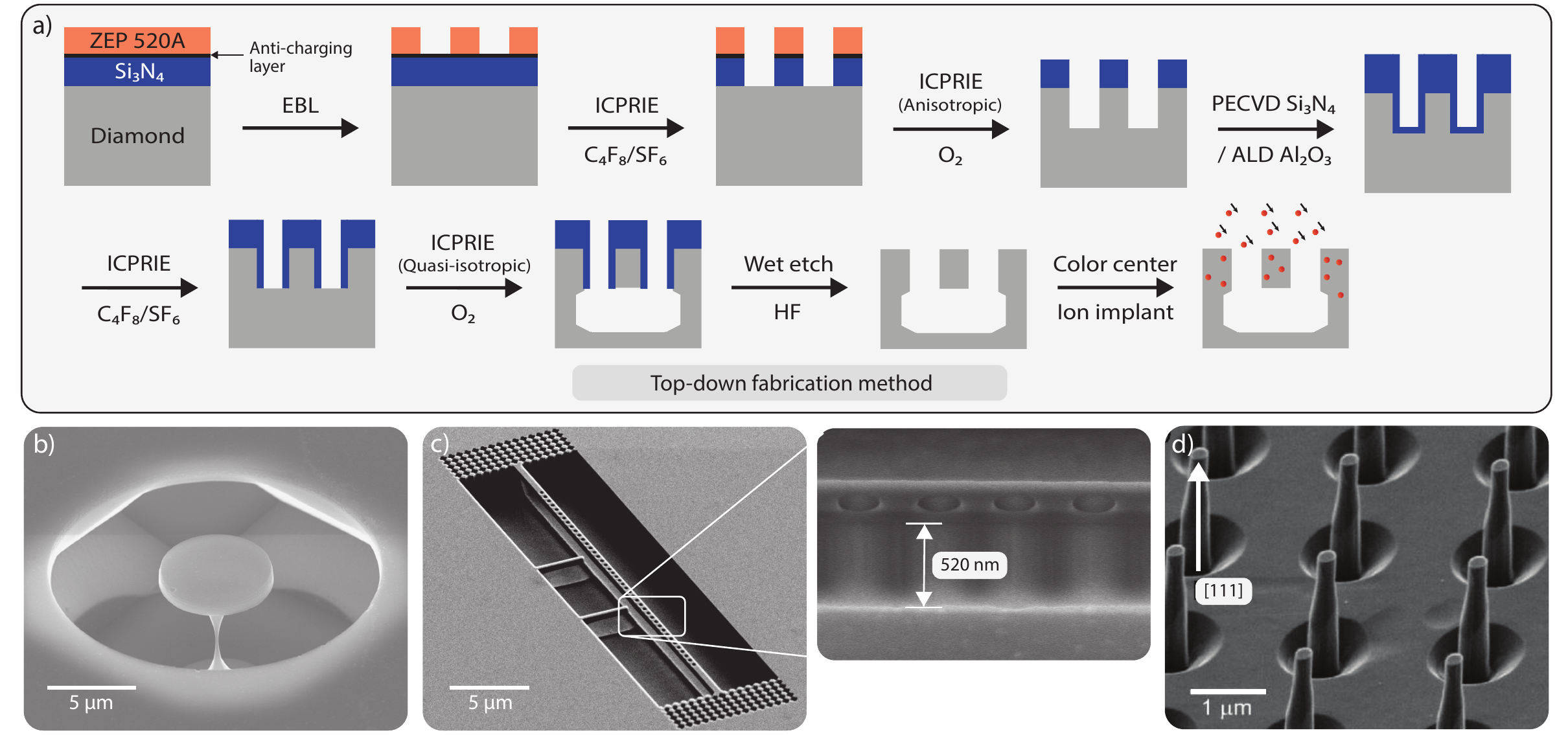}
		\caption{\textbf{Single-crystal diamond nanostructuring:} \textbf{(a)}  Process flow of nanofabrication via quasi-isotropic etch undercut.  The resist, which sits on the top of the anti-charging layer, silicon nitride (Si$_3$N$_4$) hard mask, and single-crystal diamond (SCD) stack is exposed using electron beam lithography (EBL), such that the pattern can be transferred through the hard mask via inductively coupled plasma reactive ion etch (ICPRIE).  An oxygen ICPRIE further defines the pattern into diamond.  A sidewall protection hard mask consisting of Si$_3$N$_4$ through plasma enhanced chemical vapour deposition (PECVD) or alumina (Al$_2$O$_3$) by atomic layer deposition (ALD) is anisotropically etched, removing the bottom surface. A quasi-isotropic oxygen etch undercuts the device, releasing the optomechanical device.  The residual hard mask is removed using hydrofluoric acid (HF). Color centers can be implanted in the device through methods such as localized ion irradiation, and vacancy diffusion can be promoted through high temperature annealing. Representative devices fabricated through this method are shown in \textbf{(b)} as a microdisk capable of containing whispering gallery optical modes coupled to radial and wine glass-like mechanical modes and in \textbf{(c)} as a nanomechanical beam with an integrated 1D photonic crystal cavity defined by air holes through the beam thickness (inset). \textbf{(d)} An array of diamond nanopillars defined on a $\langle111\rangle$-oriented SCD substrate substrate via ICPRIE etch using EBL-defined cylindrical etch masks. \textit{Panel (d) is reprinted from E. Neu et al., Appl. Phys. Lett. \textbf{104}, 153108 (2014), with the permission of AIP Publishing.}}
		\label{fig:Dfab}
\end{figure*}

Perhaps the most challenging aspect of diamond nanostructuring lies in identifying a mask material with sufficiently high selectivity to oxygen-based plasma etching, in order to withstand  pattern transfer into the substrate.  Polymer resists commonly used in electron beam (or photo-) lithography are not sufficient, and therefore an additional hard mask layer is required.   A common choice for a hard mask is silicon nitride (Si$_3$N$_4$) deposited by plasma-enhanced chemical vapor deposition (PECVD) with a thickness of about 250$\thinspace$nm  for anisotropic etches into diamond of several micrometers\cite{Khanaliloo2015, Mouradian2017, Faraon2011, Mitchell-2019-APLphotonics-DiamondMicrodisks,Khanaliloo2015PRX}. The process flow is schematically outlined in Fig.\,\ref{fig:Dfab}\,(a). After hard mask deposition, a thin ($\sim$5$\thinspace$nm) conductive film is deposited onto the substrate in order to provide a charge compensating layer during the electron beam lithography (EBL). This is required due to the highly insulating nature of diamond.  There are several options for use as the anti-charging layer including evaporated amorphous carbon, titanium, and niobium, with the latter two providing secondary functionality as an adhesion layer between diamond and the resist\cite{Chia2022}.  After spin-coating and baking the resist, patterning is performed with EBL followed by resist development.  In the case of positive-tone resists (such as PMMA or ZEP 520A) where scission of polymer chains occurs during development, cooling of the developer to around -10$^\circ$C will yield improved pattern and line edge resolution\cite{Ocola2006}. The pattern in the developed resist is transferred to the nitride layer using an inductively coupled plasma reactive ion etch (ICPRIE) with C$_4$F$_8$\,/\,SF$_6$ chemistry, followed by an anisotropic oxygen plasma etch to transfer the pattern to the diamond substrate. Alternatively, negative-tone HSQ electron beam resists can be patterned directly on the diamond. Patterned HSQ has an oxide-like nature that is resistant to oxygen plasma etching\cite{Chia2022}.

Quasi-isotropic diamond undercut etching can then be used to create suspended diamond devices. This technique, which was inspired by the silicon SCREAM fabrication process\cite{ref:shaw1994scr}, has been adopted in numerous studies\cite{Mouradian2017, xie2018crystallographic, Mitchell-2019-APLphotonics-DiamondMicrodisks, Rugar2021, Kuruma2021}. It requires that the sidewalls of the diamond nanostructure to be protected.  A conformal layer of silicon nitride deposited using a similar procedure as in the application of the initial hard mask, or an Al$_2$O$_3$ layer created by atomic layer deposition\cite{wan2018two}, has been used for sidewall shielding.  The horizontal surfaces of the protective layer are removed preferentially using an anisotropic ICPRIE etch, leaving the sidewalls covered while exposing the diamond surfaces to be etched.  The undercutting of the devices uses a zero DC-bias quasi-isotropic oxygen etch\cite{Khanaliloo2015} at elevated temperatures of 200-300$^\circ$C to increase the vertical and horizontal etch rates, which are dependent on the diamond crystal facet orientations in the bulk substrate.  Finally, the residual resist and the hard mask are removed in a hydrofluoric acid wet etch. Representative devices fabricated using this process in the geometries of a microdisk and one-dimensional photonic crystal cavity beam are shown in Figs.\,\ref{fig:Dfab}\,(b) and (c), respectively. 

Two alternative approaches for creating free-standing structures from bulk single-crystal diamond are reactive ion beam angled etching (RIBAE)\cite{Atikian2017,Chia2022} and Faraday cage angled reactive ion etching\cite{Burek2012}. In RIBAE, an incident beam normal to the substrate surface defines the depth of the nanostructure through a lithographically patterned mask. The substrate is then tilted, such that the ion beam is at an acute angle to the substrate surface and etching occurs under rotation, yielding suspended nanostructures with a triangular cross-section. In Faraday cage angled etching, a Faraday cage surrounding the sample is used to redirect the RF field driven ions in the plasma.

\subsection{Color Center Fabrication}
\label{Sec:Defect_fabrication}
A common feature for the color centers discussed in Sec.\,\ref{sec:ColorCenters} is the presence of an impurity atom combined with adjacent lattice vacancies. However, for many quantum applications, a high-quality diamond with a low concentration of impurities is desired to mitigate electric- and magnetic noise caused by impurity ions. Both the impurity atom and the vacancies can be introduced in various ways. For example, vacancies can be introduced by irradiation of high-energy particles, such as electrons, neutrons, and ions, or via intense, ultrafast laser pulses (see the last panel, Fig.\,\ref{fig:Dfab}\,(a))\cite{Waldermann2007, Pezzagna2011, Orwa2011,Smith2019}. High-temperature annealing ($T\ge 800^\circ\,\text{C}$) provides sufficient thermal activation energy for the vacancies to diffuse. During this diffusion process, the vacancies can combine with impurity atoms, forming color centers. The annealing is usually carried out in a high vacuum or in an inert atmosphere to avoid the etching of the diamond surface and the formation of a graphitic surface layer\cite{Chu2014NanoLett}. The bombardment of particles during irradiation leads to crystal damage--a potential source of paramagnetic noise adversely affecting the spin and optical properties of the color centers. In principle, annealing mitigates these noise sources by repairing the crystal damage\cite{Naydenov2010}. 

For the integration of color centers in devices such as photonic cavities\cite{Faraon2011,Riedel2017} and all-diamond scanning probes\cite{Maletinsky2012}, precise control of the lateral and vertical position of the color centers with respect to the relevant fields involved is of paramount importance. 
In principle, this can be accomplished by deterministically fabricating the devices around the color centers--a challenging task with the current state-of-the-art fabrication techniques. Another way to remedy this problem is to introduce impurity ions via $\delta$-doping. This technique has been successfully demonstrated for nitrogen-vacancy centers by introducing nitrogen gas during the CVD diamond growth process\cite{Ohno2012,Aharonovich2014}. The depth of the doping layer is controlled by the subsequent diamond overgrowth\cite{Ohno2012,meynell2020engineering}. The aforementioned irradiation and thermal annealing techniques can then be used to form NV centers in $\delta$-doped diamond. 

Contrary to nitrogen, silicon is rarely found in natural diamond\cite{Aharonovich2014}. However, silicon doping is readily achievable during diamond growth, for example by the incorporation of SiO$_2$ or SiC into the growth chamber\cite{Rogers2014} or by introducing silane (SiH$_4$) gas during the growth process\cite{Bolshakov2015,Sedov2017}. Similarly, the introduction of germane (GeH$_4$) gas allows the formation of GeV centers\cite{Sedov2018}.

Alternatively, impurity ions can be incorporated into the diamond post-growth via ion implantation. Here, the implantation aids the formation of vacancies, which, during thermal annealing, can form color centers\cite{Smith2019}. Ion implantation has successfully demonstrated the creation of NV centers\cite{Meijer2005,Haque2017}, SiV centers\cite{Wang-JPhB-2006-SiVsinglePhotons,Evans2016}, GeV centers\cite{Iwasaki2015}, SnV centers\cite{Tchernij2017,Gorlitz2020} and PbV centers\cite{Tchernij2018,Trusheim-PRB-2019-PbV}. However, the large size of the heavy group-IV atoms leads to greater lattice damage and larger strain in the crystal after thermal treatment\cite{Iwasaki2017}. However, shallow implantation and subsequent diamond overgrowth\cite{Rugar2020NanoLett} allow for the creation of deep group-IV defects while maintaining a crystalline environment. 

For ion-implanted samples, spatial positioning can, in principle, be achieved by implanting through an AFM tip\cite{Pezzagna2010,Riedrich-Moller2015}, by the use of lithographically defined masks\cite{Toyli2010,Sangtawesin2014} or using focused ion beam (FIB)\cite{Tamura2014,Schroder2017}. Furthermore, the ion implantation energy can be adjusted according to the desired target depth, allowing for the creation of color centers at a depth ranging from a few to several tens of nanometers. In particular, the use of FIB allows for high-precision implantation into pre-fabricated photonic structures, such as photonic crystals\cite{Schroder2017OpticalMaterials} and nanobeam cavities\cite{Sipahigil-Science-2016-SiVplatform,Evans2018,Schroder2017}. However, the creation of deeper color centers requires the use of high-energy ions. These ions lose energy in collision with electrons and atomic nuclei in the lattice, thereby leading to a trail of lattice damage along the trajectory\cite{vanDam2019}. Furthermore, the collision with nuclei causes deviation from the designated path, thereby reducing the spatial accuracy.

Vacancy generation using tightly focused, ultrafast laser pulses constitutes a promising alternative to ion implantation for the creation of NV centers deep in the diamond\cite{Chen2017NatPhot,eaton2019quantum}. In this process, vacancies are formed as a consequence of optical breakdown caused by tunnelling or multiphoton absorption\cite{Schaffer2001}. The highly nonlinear nature of this process confines the lattice damage within the focal volume of the excitation laser\cite{Chen2019Optica}. Relying on the natural occurrence of nitrogen in electronic grade diamond ($[\textrm{N}]<5\,\textrm{ppb}$), laser writing and subsequent annealing has led to the formation of highly stable NV centers\cite{Chen2017NatPhot,Stephen2019,Yurgens2021}. The creation of stable SiV centers has been demonstrated by coating the diamond surface with silicon nanoballs followed by fs laser illumination\cite{Rong2019}.

For quantum sensing applications (see Sec. \ref{Sec:Sensing}), it is necessary to have the probe spin in close proximity to the target spins\cite{grinolds2014subnanometre, rugar2015proton, haberle2015nanoscale}. The reason is twofold: firstly, magnetic dipolar coupling of probe spin with the nearby target spins scales as $1/r^3$, where $r$ is the probe-to-target spin distance. Consequently, only $r$ in the nanometer range can provide sufficient coupling strengths. Secondly, for imaging applications the spatial resolution is determined by the  minimum $r$, thus demanding close proximity of the probe and target spins, which is even more critical since the targets often lie exterior to the diamond surface. Clearly, only probe spins located a few nanometers below the diamond surface are useful for such applications. However, these so-called shallow spins are also extremely sensitive to various unwanted noise sources present on the surface, thereby severely degrading their spectral and spin properties\cite{ofori2012spin, myers2014probing, Chakravarthi2021}. In this regard, engineering robust sensing qubits is an ongoing endeavour.

\section{Quantum Photonic Devices}
\label{sec:nanophotonicdevices}

Efficient collection of photoluminescence (PL) is of paramount importance for applications using color centers in diamond. However, the large refractive index of diamond ($n=2.4$) leads to total internal reflection at the diamond-air interface, limiting the detection efficiency to only a few $\%$ for color centers in bulk diamond when using a high numerical aperture microscope\cite{Hadden2010}. To remedy this problem and improve the photon flux, a variety of different micro- and nano-photonic devices have been fabricated on diamond\cite{rani2020recent}. 

\subsection{Efficient Photon Extraction via Device Fabrication}
A widely adopted approach for improving the collection efficiency of broadband PL from color center is fabricating solid immersion lenses (SILs, see Fig.\,\ref{fig:nanophotonics}\,(a))\cite{Hadden2010,Jamali2014}. PL emitted from a color center located in the focus of the SIL will strike perpendicular to the diamond surface, thus suppressing total internal reflection. SILs have been used to demonstrate enhanced detection efficiency from NV centers\cite{Robledo2011,Marseglia2011}, SiV centers\cite{Rogers2014,Hepp2014} and GeV centers\cite{Siyushev2017,Chen2019PRL}. One key advantage of SILs is their relatively easy fabrication using FIB milling around pre-characterized color centers located deep in the diamond. This is of particular importance for experiments using NV centers, where the  proximity to surfaces often deteriorates their optical coherence\cite{Chakravarthi2021}. State-of-the-art demonstrations of remote entanglement of NV centers\cite{Bernien2013,Hensen-Nature-2015-LoopHoleFreeBell1p3km} utilized diamond SILs fabricated around naturally occurring NV centers\cite{vanDam2019}. Diamond SILs provide a detection efficiency of $\sim10\,\%$\cite{Ruf2021JAP}, which can be further improved by the application of an anti-reflective coating\cite{pfaff2014unconditional}.

Although diamond SILs suppress total internal reflection, the photon detection efficiency is still limited by the non-directional emission of the color center PL. Recently, it has been demonstrated that the detection efficiency can be further increased by using layered dielectric optical antennas\cite{Lee2011NatPhot}. Such a device combines layers of different dielectric materials, where the contrast in refractive index modifies the emission pattern, leading to preferential photon emission into the high-index material\cite{Luan2006}. In the context of diamond photonics, a low-loss, broadband dielectric antenna has been demonstrated using a diamond micromembrane ($n=2.4$) bonded to a macroscopic GaP SIL ($n=3.3$) suspended in air ($n=1.0$)\cite{Riedel2014}. As a consequence of the asymmetric refractive index profile, PL is predominantly emitted into the GaP. The hemispherical shape of the SIL suppresses total internal refraction, and can thus, in principle, provide a near-unity collection efficiency\cite{Chen2011,Chu2014Optica}. An alternative approach constitutes the use of inverse design\cite{Dory2019NatCom} to fabricate photonic structures. Recently, this approach demonstrated a 14-fold enhancement of broadband PL from NV centers located at a depth of $\sim$100\,nm\cite{Chakravarthi2020Optica}.

Nanophotonic waveguiding structures, such as nanowires\cite{Babinec2010, Hausmann2010, Marseglia2018} and nanopillars (see Fig.\,\ref{fig:Dfab}\,(d))\cite{Neu2014,Momenzadeh2015}, offer an alternative method to improve the collection efficiency of broadband PL. These devices act like antennas, directing the emitted PL into a well-defined guided mode\cite{Schroder2016}. Detection efficiency of $40\,\%$ has been demonstrated using top-down fabricated nanowires on diamond with ingrown NV centers\cite{Babinec2010}. One particular advantage of nanopillars is the possibility to embed single color centers close to the tip ($\sim10\,\textrm{nm}$) by shallow ion implantation (see Section\,\ref{Sec:Defect_fabrication}) prior to fabrication, while maintaining long spin coherence times\cite{Maletinsky2012,Appel2016}. This is of particular importance for applications in quantum sensing, where the spatial resolution depends on the sensor to sample distance\cite{Rondin2014}. Tailoring the pillar geometry can further increase the collection efficiency. Recently, a collection efficiency of $57\,\%$ was demonstrated for an NV center located at the focus of a truncated parabolic reflector diamond nanopillar\cite{Hedrich2020}.

\subsection{Engineering of the Photonic Environment}
The devices discussed so far enhance the collection efficiency of broadband PL, which is particularly useful for applications in quantum sensing, where an increased photon flux leads to improved sensitivity to magnetic fields\cite{Rondin2014}. 
However, quantum information applications require a high flux of coherent, indistinguishable photons, and therefore modification of the optical properties of the color centers is often desired. The rate of remote entanglement in schemes relying on two-photon quantum interference is limited by the detection rate of coherent photons. At the time of writing, the current state-of-the-art experiment using NV centers achieved an entanglement rate of $\sim10\,\textrm{Hz}$\cite{Humphreys-Nature-2018-RemoteEntanglement,pompili2021realization}, limited by the long radiative lifetime ($\tau\simeq12\,\textrm{ns}$), the aforementioned small branching into the ZPL (Debye-Waller factor of $\sim3\,\%$) and poor photon extraction efficiency owing to total internal reflection. While the group-IV defect centers possess more favorable optical properties in terms of a larger Debye-Waller factor and shorter radiative lifetime, to date, no experiments demonstrating remote entanglement have been conducted, though recently indistinguishable photons from SnV color centers were reported\cite{arjona2022photonic}.

\begin{figure}[t!]
	\includegraphics[width=\linewidth]{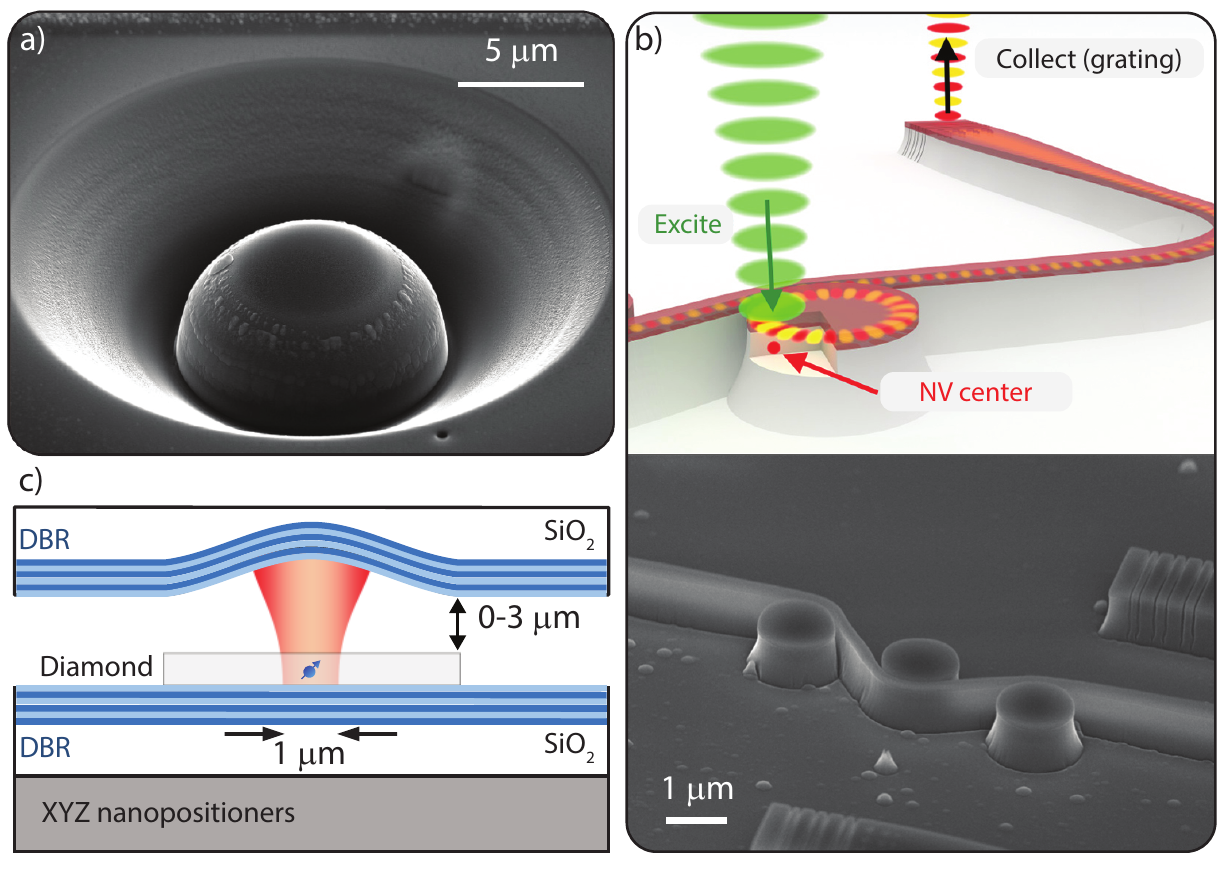}
		\caption{\textbf{Photonic devices for efficient light extraction:} \textbf{(a)} SEM image of a SIL fabricated using FIB milling.
		\textbf{(b)} Top panel: schematic of a hybrid photonic platform based on GaP on diamond. Evanescent coupling between color centers and optical resonators enhances the photon emission, which is further waveguided and detected via grating out-couplers. Bottom panel: SEM image of the hybrid photonic device.
		\textbf{(c)} Schematic of a diamond membrane embedded in an open microcavity. The fully tunable platform allows for \textit{in situ} tuning of both the cavity resonant frequency and the position of the color center with respect to the cavity mode.
		\textit{Panel (a) is reprinted from M. Jamali et al., Rev. Sci. Instrum. \textbf{85}, 123703 (2014) with permission of AIP Publishing.
		Panel (b) is reprinted with permission from M. Gould et al., Phys. Rev. Applied \textbf{6}, 011001 (2016), Copyright 2016 by the American Physical Society.
		Panel (c) is reprinted with permission from D. Riedel et al., Phys. Rev. X \textbf{7}, 031040 (2017) licensed under the terms of the Creative Commons Attribution 4.0 license.}
		}
		\label{fig:nanophotonics}
\end{figure}

In principle, these shortcomings can be addressed by embedding color centers in photonic resonators\cite{Su2008}. Resonant coupling of the ZPL to a single cavity mode enhances the ZPL emission on two grounds. First, the cavity directs the emission into a well-defined output mode, improving the photon collection efficiency\cite{Gao2015,Flagan2022}. Second, the emitter experiences a Purcell effect\cite{Purcell1946} that enhances the spontaneous emission rate of transitions resonant with a cavity mode. This can be used to increase the fraction of photons emitted into the ZPL\cite{Riedel2017}. For an optical cavity with quality factor $Q$ and mode volume $V$, the Purcell factor is given by
\begin{equation}
\begin{split}
   F_{\textrm{P}}&=F_{\textrm{P}}^{\textrm{max}}\xi\frac{1}{1+4Q^2\left(\lambda_{\textrm{ZPL}}\,/\lambda_{\textrm{cav}}-1\right)^2}\,, \\
    F_{\textrm{P}}^{\textrm{max}} &= 1+\frac{3}{4\pi^2}\frac{Q}{V}\left(\frac{\lambda}{n}\right)^3\,, \\
\end{split}
\end{equation}
where $\xi=\left(\frac{\abs{\boldsymbol\mu\cdot \mathbf{E}}}{\abs{\boldsymbol\mu}\cdot\abs{\mathbf{E}}}\right)^2$ describes the overlap between the dipole moment $\boldsymbol\mu$ and the electric field $\mathbf{E}$. Note that  $F_{\textrm{P}}$ is independent of any emitter properties--$F_{\textrm{P}}$ is described solely by the characteristics of the cavity\cite{Flagan2022}. The scaling $F_{\textrm{P}}\propto Q\,/V$ motivates the use of high-quality resonators with minimal mode volume\cite{Janitz2020}.

Coupling between color centers in diamond and photonic resonators has been demonstrated on various different platforms\cite{Janitz2020}, including, but not limited to, photonic crystal cavities\cite{Englund2010,Faraon2012,Riedrich-Moller2012,Riedrich-Moller2014,Jung2019}, nanobeam cavities\cite{Li2014,Sun2018PRL,Evans2018,Nguyen2019,bhaskar2020experimental,Rugar2021,Kuruma2021}, waveguides\cite{Bhaskar2017a,Hadden2018,Rugar2020}, microrings\cite{Faraon2011,Hausmann2013,Faraon2013}, hybrid optical devices\cite{Barclay2011,Gould2016,Gould2016JOSAB,Zhang2016,Wan2020,Fehler2021} and open microcavities\cite{Johnson2015,Kaupp2016,Riedel2017,Benedikter2017,Haussler2019,Jensen2020,Ruf2021}. 

In photonic crystal cavities (PCC), the periodic change in refractive index creates a photonic bandgap. Tailoring of the periodicity can confine light to a mode volume of $\lesssim(\lambda\,/n)^3$. However, at the visible wavelengths relevant for color centers in diamond, imperfection in the fabrication of these devices limits the achievable $Q$-factor to $\sim10^3-10^4$. The current state-of-the art PCC containing SiV centers exhibits $Q=2\times10^4$ and $V=0.5(\lambda\,/n)^3$\cite{bhaskar2020experimental}. Note that PCCs with $Q>10^5$ have been demonstrated for $\lambda=1,550\,\textrm{nm}$\cite{Burek2014}. As discussed in Section\,\ref{Sec:Defect_fabrication}, deterministic coupling of color centers to PCCs can be achieved via FIB milling around pre-characterized color centers\cite{Riedrich-Moller2014}, or by using a FIB to implant ions in fabricated structures. The latter approach has been proven to be particularly successful for SiV centers\cite{Evans2018}. Furthermore, red-shifting of the cavity resonance is possible via deposition of $\textrm{N}_{2}$\cite{Evans2018} or Ar\cite{Rugar2021} gas, thereby maximizing the spectral overlap between the color canter ZPL and the cavity mode. 

Coupling to WGMs in diamond microdisk- and ring resonators offer a monolithic alliterative to PCCs. In general, WGMs offer a larger $Q$-factor, albeit at the expense of a larger $V\sim10(\lambda\,/n)^3$. In practice, due to fabrication imperfection and surface roughness, the achievable $Q$-factors are comparable to those of PCCs\cite{Faraon2011}. Nevertheless, due to advances in device fabrication, a diamond microdisk with $Q=11,000$ has been demonstrated for $\lambda=638\,\textrm{nm}$\cite{Mitchell-DiamondOptomechanicalResonator-Optica-2016}.

The monolithic photonic devices discussed above offer large $Q\,/V$ ratios at the expense of invasive nanofabrication. While NV centers with close-to lifetime limited optical linewidths have been reported in bulk\cite{Tamarat2006,Chu2014NanoLett}, NV centers embedded in nanophotonic devices often suffer from poor optical coherence and inhomogeneous broadening of the ZPL linewidth on the account of a fluctuating charge environment caused by fabrication induced surface damage\cite{Kasperczyk2020,Chakravarthi2021}. Therefore, increasing the defect free, crystalline environment has proved to be beneficial\cite{Ruf-NanoLett-2019-CoherentNVinMembrane}.

Hybrid photonic platforms constitute an alternative to monolithic resonators on the account of the possibility to use bulk diamond. These devices (see Fig.\,\ref{fig:nanophotonics}\,(b)) combine single-crystal diamond with waveguides and resonators fabricated from a high-index material\cite{Barclay2009, thomas2014waveguide}. Evanescent coupling of NV centers to waveguides\cite{Fu2008,Gould2016} and microcavites\cite{Barclay2011,Fu2011} has been demonstrated on various platforms based on GaP\cite{Gould2016JOSAB}. However, while red-tuning of the cavity resonance are possible via gas deposition, one remaining drawback of these hybrid devices is the difficulty in positioning the color center close to the field maxima of the guided mode\cite{Gould2016}. Furthermore, evanescent coupling requires near-surface NV centers, which, as discussed above, suffer from inhomogenous linewidth broadening\cite{Chakravarthi2021}. Nevertheless, hybrid photonic devices offer a viable route to integrate color centers in diamond with large-scale photonic integrated circuits\cite{Mouradian2015,Schmidgall2018}. The state-of-the-art experiment demonstrated a 128-channel `quantum microchiplet', a diamond waveguide array containing highly-coherent SiV and GeV centers interfaced with an integrated AlN photonic circuit\cite{Wan2020}.

In recent years, planar-concave open Fabry-Perot microcavities (see Fig.\,\ref{fig:nanophotonics}\,(c)) have emerged as a compelling alternative to the monolithic and hybrid optical resonators discussed above\cite{Riedel2017,Jensen2020,Ruf2021}. These cavities are formed from a highly-reflective planar distributed Bragg-reflector (DBR), to which a diamond membrane is bonded. A second highly-reflective DBR mirror, with micron-sized concave indentations fabricated using FIB milling\cite{Dolan2010,Johnson2015} or CO$_{2}$ laser ablation\cite{Hunger2012,Greuter2014,Najer2017}, concludes the cavity. The Gaussian-shaped indentations facilitate efficient coupling of the cavity mode to single-mode external detection optics\cite{Riedel2020,Tomm2021}. The open microcavity offers the possibility to incorporate micron-sized single-crystal diamond membranes\cite{Janitz2015}, thus preserving the optical coherence of NV centers\cite{Ruf2021}, while maintaining a large $Q\,/V$-ratio\cite{Flagan2022}. Furthermore, with the use of piezoelectric nanopositioners, the open microcavity platform offers full $\textit{in situ}$ tunability and control of both the cavity resonant frequency and the relative position of the color center with respect to the cavity mode\cite{Riedel2017}. Resonant coupling of a single NV center to an open microcavity has demonstrated enhancement of the fraction of ZPL photons to $\sim46\,\%$\cite{Riedel2017}. Finally, the incorporation of a diamond membrane with a small thickness gradient has demonstrated  full \textit{in situ} control of both the resonant frequency, and the relative frequency spacing of adjacent cavity modes\cite{Flagan2021Dres}, thus providing a platform for tunable nonlinear optics.

\section{Sensing and Metrology}
\label{Sec:Sensing}

Quantum sensing and metrology involve the detection and measurement of physical quantities using  quantum properties such as coherence and entanglement, with the goal of reaching fundamental limits in the measurement\cite{giovannetti2004quantum,Degen-QuantumSensing-RMP-2017}. Varieties of such sensors have been developed in the past several decades\cite{Degen-QuantumSensing-RMP-2017}. Among the solid-state platforms, spins in diamond have emerged as versatile sensors for a variety of physical quantities, both classical and quantum\cite{schirhagl2014nitrogen}. The NV center outperforms other emitters in diamond primarily due to its excellent spin (e.g.\ long coherence times) and optical (e.g.\ photostability, brightness) properties. 

\subsection{NV Center-Based Sensing}
\begin{figure}[t!]
	\includegraphics[width=\linewidth]{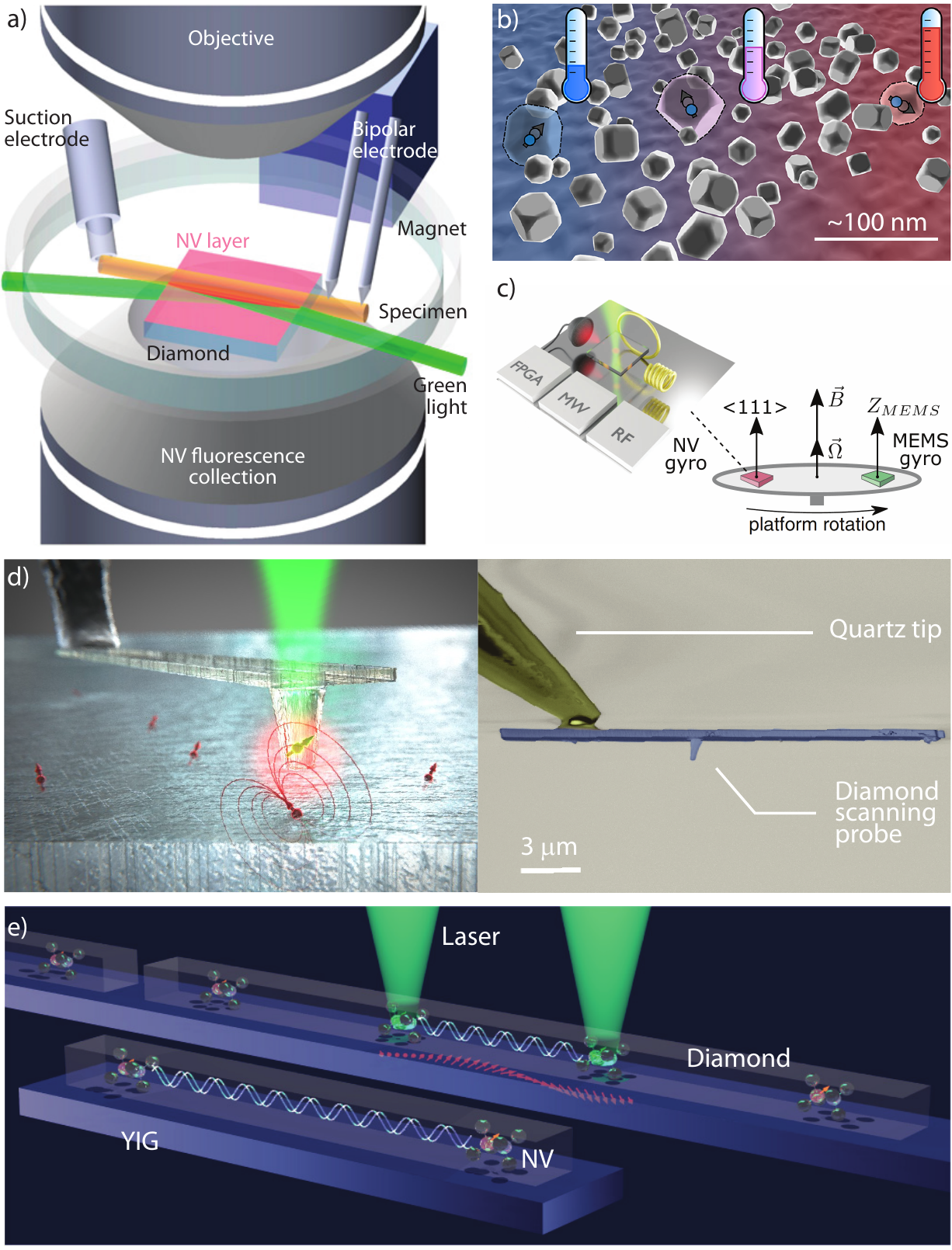}
		\caption{\textbf{Diamond-based quantum sensing with spins as multi-modal sensors: }\textbf{(a)} Diamond NV centers for sensing biomagentic fields, like those from, for example, neuronal action potential from an organism at ambient conditions. \textbf{(b)} NV centers as temperature sensors with high sensitivity (mK\,/\,$\sqrt{\text{Hz}}$) and spatial resolution\cite{neumann2013high}.  \textbf{(c)} Realization of NV spins as rotation sensor with an aim of miniaturized high precision quantum gyroscope. The schematic is of an experiment that compares sensor performance with a standard rotation sensor\cite{soshenko2021nuclear}. \textbf{(d)} All-diamond nanopillar housing a single NV defect as the scanning probe for high spatial resolution magnetic imaging. Left panel, green and red arrows indicate NV and target spins, respectively. Right panel is the SEM image of the scanning probe. \textbf{(e)} NV centers in diamond on top of an infinitely long magnon waveguide for magnon-mediated spin-spin entanglement. \textit{Panel (a) is reprinted with permission from J. F. Barry et  al., PNAS, 113(49)14133-14138, 2016. Panel (b) is reprinted with permission from P. Neumann et al., Nano Lett. \textbf{2013} 13 (6), 2738-2742. Copyright 2013 American Chemical Society. Panel (c) reprinted with permission from V. V. Soshenko et al., Phys. Rev. Lett. \textbf{126}, 197702 (2021), Copyright 2021 by the American Physical Society. Panel (d, left) is adapted with permission from P. Maletinsky, Quantum Sensing Lab, University of Basel. Panel (d, right) is reprinted from P. Appel et al., Rev. Sci. Instrum. \textbf{87}, 063709 (2016), with the permission of AIP Publishing. Panel (e) is reprinted with permission from M. Fukami et al., Phys. Rev. X Quantum \textbf{2}, 040314 (2021) licensed under the terms of the Creative Commons Attribution 4.0 license.}
		}
		\label{fig:sensing}
\end{figure}

The NV center's electron spin has proven to be an exquisite probe for measuring several physical quantities, both external and internal to the host diamond, like electric and magnetic fields\cite{Degen-QuantumSensing-RMP-2017, Barry-RMP-2020-NVmagnetometry}, temperature\cite{acosta2010temperature,neumann2013high}, pressure\cite{ho2021recent}, strain\cite{trusheim2016wide}, rotation\cite{soshenko2021nuclear}, and charge\cite{dolde2014nanoscale} with high sensitivity (see Fig.\,\ref{fig:sensing}). Diamond based sensing has been intensely pursued and is rapidly progressing owing to the potentially immediate commercial and fundamental interests in physics, chemistry, biology and clinical research\cite{schirhagl2014nitrogen,Degen-QuantumSensing-RMP-2017,Barry-RMP-2020-NVmagnetometry}.

One added advantage of the NV spin as a sensor is its atomic size, which provides extremely high spatial resolution, outperforming any other sensors developed to date\cite{grinolds2014subnanometre}. Depending on the application in question, NV-based sensors have been realized using ensembles or single NV centers in diamond nano-, micro-, or bulk crystals\cite{schirhagl2014nitrogen, Barry-RMP-2020-NVmagnetometry}. One area where the NV sensors have made a profound impact is the realization of nanoscale nuclear magnetic resonance (NMR) spectroscopy\cite{glenn2018high} and magnetic resonance imaging (MRI) at ambient conditions and moderate bias magnetic fields\cite{rugar2015proton,haberle2015nanoscale}. NMR using NV centers has addressed the concerns of the low sensitivities associated with traditional NMR methods\cite{glover2002limits} and the complexities associated with advanced methods like magnetic resonance force microscopy (MRFM)\cite{degen2009nanoscale}. Recent advances in NV nano-NMR methods have placed NV centers at the forefront of molecular-scale NMR with unprecedented spectroscopic\cite{glenn2018high} and structural resolution\cite{abobeih2019atomic}.

For magnetic resonance based methods to work at the level of a single target molecule or atom, the probe has to have sufficient sensitivity to individual spins as well as possess fine spectral resolution. In this regard, NV centers have shown to be sensitive to individual molecules and spins external to the diamond surface. An NV spin, being an atomic scale probe, is capable of measuring NMR signals from target spins in about $(\text{5}\,\textrm{nm})^3$ sample volumes\cite{mamin2013nanoscale,staudacher2013nuclear} with a sub-Angstrom spatial resolution\cite{zopes2018three,abobeih2019atomic} at ambient conditions. In comparison, traditional NMR requires nanoliter sample volumes and the spatial resolution is limited to $\mu$m-scale\cite{glover2002limits}, excluding their application to the level of single live cells, for instance. Recent advancements in NV sensing protocols have also allowed the achievement of finer spectroscopic details of the target sample. These approaches include utilizing long-lived nuclear spin lifetimes\cite{aslam2017nanoscale} and quantum heterodyne methods\cite{schmitt2017submillihertz,boss2017quantum,glenn2018high} to achieve several orders of magnitude improvement in spectral resolution, effectively bypassing the standard limits set by the $T_2$ and $T_1$ times of NV spin. The demonstrated sensitivities are sufficient to resolve subtle chemical signatures in picoliter sample volumes\cite{glenn2018high}. In addition, NV centers have shown to be promising sensors in condensed matter\cite{casola2018probing} and biomagnetism\cite{schirhagl2014nitrogen,barry2016optical}, and considerable attempts have been made in using NVs for detecting dark matter\cite{marshall2021directional}.

Magnetometers are ubiquitous in many areas of application. The current state-of-the-art magnetometers (e.g.\ superconducting quantum interference devices or SQUIDs and optically pumped magnetometers or OPM) suffer from non-ambient operating conditions, modest spatial resolution, and limited dynamic range, albeit offering the highest sensitivities. To this end, progress in NV-magnetometers over the last decade has proven to be promising, however, some key challenges remain to be addressed\cite{schirhagl2014nitrogen, Barry-RMP-2020-NVmagnetometry}. The primary challenge is that the sensitivities achievable with NV magnetometers are worse than both  the theoretical predictions and the competing techniques (SQUIDs and OPMs) by at least three orders of magnitude\cite{Degen-QuantumSensing-RMP-2017,Barry-RMP-2020-NVmagnetometry}. This is mainly due to inefficient NV spin readout methods and  limited coherence times. For a diamond sample with $N$ number of NV sensors per volume ($N$=1 for a single NV center), the magnetic field sensitivity, $\eta_\text{B}$, scales as\cite{Taylor2008,dreau2011avoiding}
\begin{equation}
\eta_\text{B}\propto\frac{1}{C'}\frac{1}{\sqrt{N.T'}}\,,
\end{equation} 
where $C'$ is the readout fidelity, and $T'$ is the interrogation time (here, the spin initialization and readout times are neglected). $C'$ is determined by the PL photon collection efficiency and the contrast of the spin states, both of which are far below unity, thus rendering $C'\ll$1. $\eta_\text{B}$ is limited primarily by the photon shot noise since the readout is performed optically\cite{Barry-RMP-2020-NVmagnetometry}, and ultimately by the quantum projection noise, $\eta_\text{Bp}$, so that
$\eta_\text{B}=\eta_\text{Bp}\,/C'$\cite{Degen-QuantumSensing-RMP-2017}. Several approaches have been explored to improve $C'$ beyond the limits of standard optical readout by employing non-PL based readout schemes. However, some key challenges remain for all these methods in terms of optimizing the overhead (spin initialization and readout) times, added experimental complexities, and limited readout fidelities.  
As an alternative to PL detection schemes, cavity-enhanced readout methods, based on the IR absorption by the NV center singlet state\cite{chatzidrosos2017miniature} and interaction between NV and microwave photons in a resonator\cite{eisenach2021cavity} have been demonstrated. While these methods have substantially enhanced the fidelity $C'$ (and hence $\eta_\text{B}$) by achieving near unity IR light collection efficiency\cite{chatzidrosos2017miniature} and unity contrast\cite{eisenach2021cavity}, the overall $\eta_\text{B}$ is still short of reaching the $\eta_\text{Bp}$. Further miniaturization of the sensor in these cavity-based systems for high spatial resolution remains challenging due to their mm-lengthscale. In this context, diamond microcavities could fill this gap given their sub-$\mu$m size, and appreciable $Q\,/V$ ratio (see Sec.\ref{sec:nanophotonicdevices}-B)\cite{Khanaliloo2015}, thus potentially enabling stronger interaction between spins and cavity fields, leading to better $\eta_\text{B}$. However, it is important to understand the charge state dynamics of the NV centers under extreme optical powers in microcavities as they strongly affect $C'$.

The next critical factor influencing  $\eta_\text{B}$ is $T'$: $\eta_\text{B}$  improves with longer $T'$. The nature of the signal that can be probed depends on the characteristic timescales of the sensor spins: their inhomogeneous dephasing time ($T_2^*$) and spin coherence time ($T_2$). Short-lived $T_2^*$ times allow for the detection of only static or slowly varying target $B$-fields such as those in biological samples\cite{barry2016optical} or condensed matter\cite{casola2018probing}. On the other hand, long-lived $T_2$ times can probe AC fields or fast oscillations like those from the spins inside\cite{taminiau2012detection} and outside diamond crystal\cite{mamin2013nanoscale,staudacher2013nuclear}. Engineering qubits with maximum $T_2^*$ and $T_2$ directly affects quantum sensing strategies involving phase accumulation, and is therefore currently one of the major material fabrication challenges.

The remaining parameter influencing $\eta_\text{B}$ is $N$. However, arbitrarily increasing $N$ degrades the $T_2^*$ and $T_2$ times due to enhanced dipolar interaction among the sensing spins (NV-NV) and with substitutional nitrogen spins\cite{edmonds2021characterisation}. Another drawback of sensing using a large $N$ is the compromise of a key advantage of the NV center spin: its atomic spatial resolution. Currently, the best-reported $\eta_\text{B}$ are in the range of $\upmu$T (for DC)\cite{Taylor2008} and nT (for AC)\cite{Balasubramanian2009,zhao2022sub} for a single NV, and about pT (DC and AC) for an ensemble NV sensor\cite{Barry-RMP-2020-NVmagnetometry,wolf2015subpicotesla}.  

In summary, a crucial challenge to realize diamond as a robust quantum sensor performing at its fundamental limits lies in achieving unity readout fidelities and engineering high-quality, affordable, diamond samples with prolonged coherence times. From a material engineering point of view, a complete understanding and mitigation of the undesired surface-induced noise sources are essential for sensing target signals. Ongoing investigations on the novel color centers in diamond may overcome the challenges posed by NV centers as a sensor.

\subsection{Hybrid Diamond-Magnetic Sensors}
A related area of research that could benefit from the properties of diamond constitutes the development of nano-optomechanical torque sensors, which have demonstrated enhanced sensitivities to mechanical torque\cite{ref:wu2014ddo, ref:wu2012pcp, kim2016approaching}. They naturally respond to the fundamental interaction of magnetic moments with an external field--the magnetic torque--at a magnitude proportional to the total magnetization\cite{losby2018} providing insight into the dynamics of nanomagnetic structures.

The coupling of magnetic and mechanical systems has enabled observation of the Barkhausen effect from nanoscale magnetic defects\cite{wu2017nanocavity, hajisalem2019two} and spin resonances in nanomagnets\cite{kim2018}. A longstanding goal for these devices is the demonstration of strong coupling of magnetic torques to mechanical degrees of freedom, granting coherent information transfer between the two systems\cite{kovalev2005}.  A more ambitious target would be the realization of torque sensitivities on the order of $\sim 10^{-29} {\rm N m / \sqrt{Hz}}$, corresponding to the torque generated by a single Bohr magneton under a field of 1 A\,/\,m. In some ways, analogous to single-spin detection achieved nearly two decades ago using magnetic force microscopy\cite{ref:rugar2004ssd}, the observation of pure magnetic torque remains elusive.

To date, diamond-based nano-optomechanical sensors of magnetic torque have yet to be realized. Their development would welcome increased torque sensitivities based on diamond's favourable mechanical and optical properties compared to previously developed silicon-based sensors (see Secs.\,\ref{Sec:cavity_om} and \,\ref{Sec:Dfab}). From a device performance point of view, while it is possible to engineer good mechanical devices in silicon at low temperatures\cite{MacCabe-UltraLongPhononicLifetime-Science-2020}, the performance of these devices is, in general, limited by coupling to surface defects; a problem that is difficult to circumvent\cite{hauer2018two}. However, diamond exhibits much lower mechanical dissipation, and it is therefore expected that the mechanical properties of diamond will surpass those of silicon. In addition, the detection sensitivities can be further enhanced by operation at high optical powers, thanks to diamond's large optical bandgap and excellent thermal properties\cite{Mildren2013ch1}. The challenge of affixing magnetic material to the sensor without detrimentally affecting its optomechanical coupling and the suppression of unwanted noise remains.

Recent diamond quantum sensing efforts are directed toward creating hybrid quantum technologies involving magnetic excitations. Magnons are quantized spin waves in magnetic materials. NV center-based magnetic resonance has been used to detect and image the stray fields generated by spin waves with nanometer-scale spatial resolution. This has enabled time-domain imaging of their coherent transport, dispersion, and interference\cite{bertelli2020} and has been shown to have the sensitivity needed for imaging  static magnetization in monolayer van der Waals materials\cite{thiel2019}. Magnons have also been proposed as mediators of quantum information\cite{tanji2009heralded, sarma2021cavity, yuan2022quantum} and as a means for quantum control and enhanced readout of spin qubits due to their strong local magnetic fields\cite{kikuchi2017}.  The long coherence length of spin waves has enabled coherent control of color centers over distances of 200$\,\mu$m\cite{andrich2017}. An appealing opportunity exists for long-range spin qubit entanglement schemes mediated by magnons (a schematic is shown in Fig.\,\ref{fig:sensing}\,(e)), with predicted cooperativities exceeding unity for NV-NV coupling at low temperature\cite{fukami2021}. 

\section{Qubit-Photon Interface}
\label{Sec:interface}

As described at the outset of this review, the realization of quantum networks requires the interconnection of remote network nodes\cite{Kimble2008}. Recently, two distant superconducting qubits housed in separate cryostats were connected using a cryogenically cooled microwave waveguide\cite{Magnard2020}. Room-temperature optical links have been used to efficiently distribute quantum states over long distances at room temperature, owing to the small interaction cross-section of photons with the environment\cite{Northup2014}. However, a major hurdle toward the realization of a large-scale quantum network is the development of efficient interfaces between stationary qubits and optical photons\cite{Borregaard2019}.

\begin{figure}[t!]
	\includegraphics[width=\linewidth]{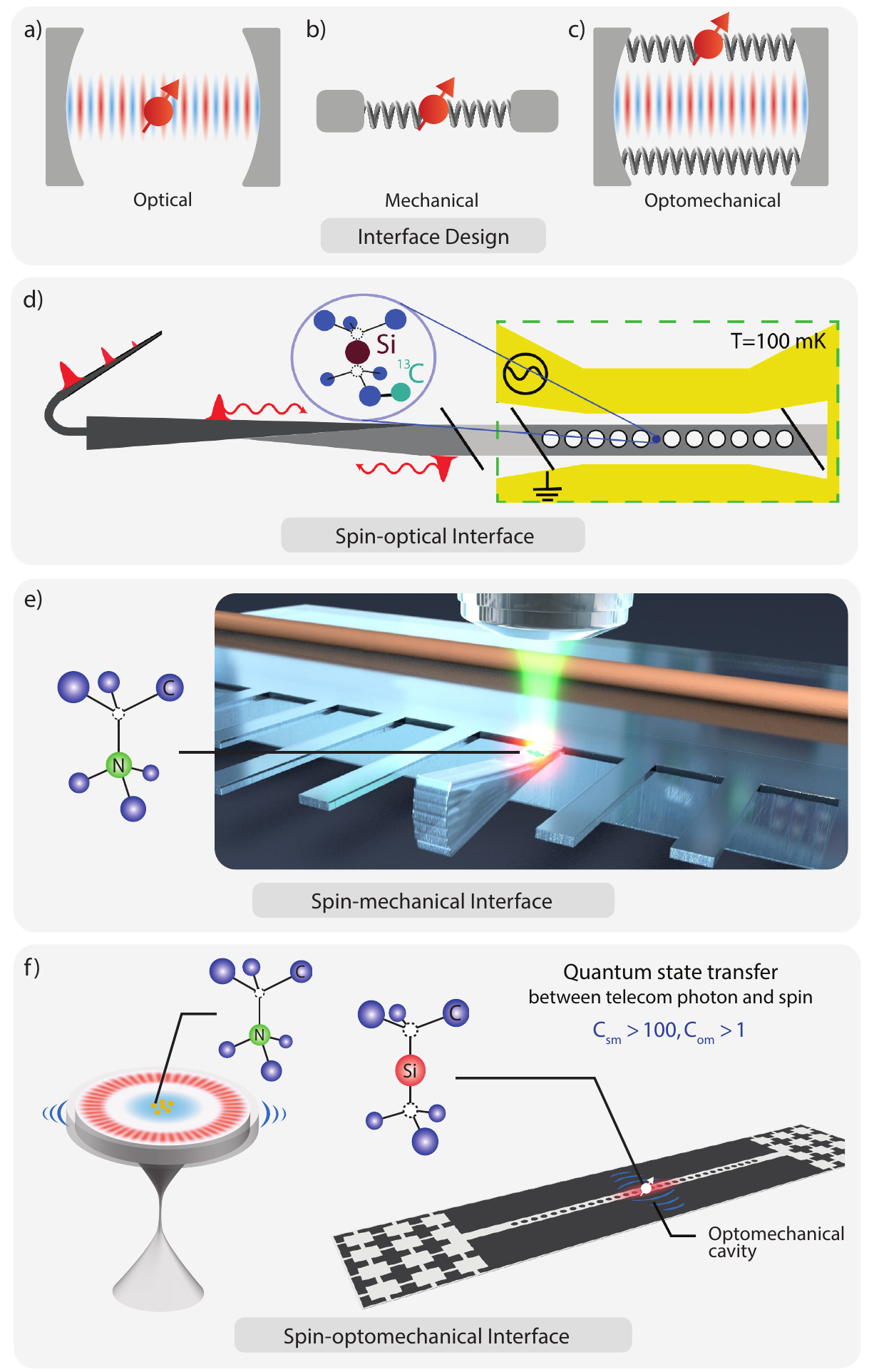}
		\caption{\textbf{Development of qubit-photon interfaces:} Cartoons of different canonical interfaces. \textbf{(a)} Qubit-photon interface based on cavity QED, \textbf{(b)} spin-mechanical interface, and \textbf{(c)} cavity optomechanical qubit-photon interface. Schematic of the experimental setup used for \textbf{(d)} a spin-photon interface using a cavity QED system\cite{bhaskar2020experimental,Nguyen2019a}, \textbf{(e)} a spin-mechanical interface as a step towards cavity optomechanical spin-photon interfaces\cite{Barfuss-StrongMechanicalDrivinfNV-NatPhys-2015}. \textbf{(f, left)} Diamond cavity optomechanical interface between spin and telecommunication photons\cite{shandilya2021optomechanical}. \textbf{(f, right)} Spin-photon interface based on optomechanical crystal cavities with phononic shields can reach $C_{\textrm{sm}}>100$ and $C_{\textrm{om}}>1$. These next-generation spin-photon interfaces can be realized with current state-of-the-art devices. 
		\textit{Panel (d) is reprinted with permission from C.T. Nguyen et al., Phys. Rev. Lett. \textbf{123}, 183602, Copyright 2019 by the American Physical Society. Panel (e) is adapted with permission from P. Maletinsky, Quantum Sensing Lab, University of Basel.}
		}
		\label{fig:interface}
\end{figure}

Coherent qubit-photon coupling is inherently limited by the weak light-matter interaction\cite{ReisererRempe-RMP-2015-CavityQED}. However, as depicted in Fig.\,\ref{fig:interface}\,(a), cavity quantum electrodynamics (QED) provides a promising route to overcome this hurdle by strong confinement of light inside optical resonators, thus enhancing the qubit-photon interactions\cite{Najer-Nature-2019-StronglyCoupledQD}. In principle, solid-state emitters embedded in nanophotonic cavity QED devices offer scalable fabrication\cite{Mouradian2017APLphotonics}, on-chip photonic routing\cite{Wan2020}, and electronic- and mechanical control\cite{Anderson2019,MacHielse2019}: key requirements for integrated network nodes in large-scale quantum networks.

For quantum network applications, the spin-photon interface has to operate in a regime where a single photon coherently and reversibly couples to a spin qubit. In general, the coupling between two different quantum systems is characterized by the cooperativity parameter $C$, defined as\cite{ReisererRempe-RMP-2015-CavityQED}
\begin{equation}
C=\frac{4g^2}{\gamma_1\gamma_2}\,, \label{eq:cooperativity}
\end{equation}
where $g$ is the coupling rate between the systems and $\gamma_{1,2}$ are the energy decay rates of each system. The condition $C > 1$ enables coherent interaction between the two quantum systems, despite their internal decoherence, and has been demonstrated for a variety of systems, including, but not limited to,  atoms\cite{Boca2004}, superconducting resonators\cite{Fink2008}, molecules\cite{Wang2019NatPhys,Pscherer2021}, and semiconductor QDs\cite{Najer-Nature-2019-StronglyCoupledQD}. However, the optical coherence of solid-state emitters is strongly influenced by the host material\cite{Aharonovich-NatPhot-2016-Solid-StateSinglePhotonEmitters,Ruf2021JAP}: interactions with the local environment leading to the inhomogeneous broadening of the optical linewidth and consequently loss of photon coherence. The effect of inhomogeneous broadening (with rate $\gamma^*$) can be incorporated by introducing the coherence cooperativity\cite{Borregaard2019,Ruf2021JAP}, where $\gamma_2\rightarrow\gamma_2+\gamma^*$ in  Eq.\,\ref{eq:cooperativity}\cite{Flagan2022}.

Remote entanglement protocols relying on two-photon quantum interference require a high flux of coherent indistinguishable photons. In the Barrett and Kok protocol\cite{ref:barrett2005ehf}, successful entanglement events are heralded by the detection of two independent ZPL photons. The overall success rate of this protocol scales with $\frac{1}{2}\eta_{\textrm{ZPL}}^2$, where $\eta_{\textrm{ZPL}}$ is the probability of detecting a ZPL photon and the factor $\frac{1}{2}$ accounts for the need to detect two photons per successful entanglement event. As discussed in Section\,\ref{sec:nanophotonicdevices}, $\eta_{\textrm{ZPL}}$ can be drastically enhanced by utilizing the Purcell effect in cavity QED devices. 

To date, all experiments demonstrating remote entanglement of color centers in diamond\cite{Bernien2013, Hensen-Nature-2015-LoopHoleFreeBell1p3km} have been conducted using native NV centers and SILs\cite{vanDam2019}. Reversible spin-photon coupling requires $C_{\textrm{sp}}>1$, a condition that cannot be satisfied using SILs. In practice, NV centers embedded in cavities typically suffer from inhomogeneous linewidth broadening\cite{Riedel2017}, manifested by compromised photon indistinguishability and $C_{\textrm{sp}}\ll1$. While recent results using native NV centers in nanopillars are promising\cite{Orphal-Kobin2022}, for network nodes using NV centers, the long-term optical stability remains a significant hurdle. On the other hand, nanoscale positioning of SiV centers in nanocavities, depicted schematically in Fig.\,\ref{fig:interface}\,(d), shows remarkable properties; $C_{\textrm{sp}} > 100$\cite{bhaskar2020experimental}, enabling spin-state dependent photon reflection and single-shot spin-state readout with fidelity $> 99.9\,\%$, combined with a fiber collection efficiency surpassing $90\,\%$\cite{Burek2017} have been demonstrated.  However, at the time of writing, entanglement of remote SiV centers, or any other group-IV defects, remains to be reported.

\subsection{Strategies for a Universal Qubit–Photon Interface}
It is likely that future long-distance quantum networks will make use of existing low-loss fiber infrastructure. However, to date, most cavity-spin systems operate at wavelengths resonant with the spin optical transition, usually in the visible range where fiber loss is high relative to telecommunication wavelengths. Coherent, entanglement preserving quantum frequency down-conversion from visible to telecommunication wavelengths is therefore required to minimize losses in the fiber links\cite{Dreau-PRAppl-2018-WavelengthConversionNV,tchebotareva2019entanglement,Krutyanskiy-npjQI-2019-AtomEntangWavelengthConversion,Yu-Nature-2020-EntanglementViaFiberOverDozensKm}. However, cavity QED systems with subsequent frequency down-conversion, while promising, come with limitations. First and foremost, for this approach to work, the qubit system must exhibit well-defined intrinsic spin-conserving optical transitions, a feature absent for popular qubit platforms such as gate-defined quantum dots and superconducting resonators. Second, differences in the local environment leads to spectral wandering, rendering photon emission from two cavities distinguishable. At the time of writing, cavity-enhanced entanglement of remote color centers in diamond has yet to be demonstrated.

An active area of research to tackle these challenges at a fundamental level involves using mechanical systems. Mechanical degrees of freedom are central to many quantum technologies\cite{Barzanjeh2022}, thanks to their ability to couple to a wide range of fields--electrical, magnetic, optical, and gravitational--through device engineering\cite{treutlein2014hybrid}. Nanomechanical systems can be naturally integrated with solid-state quantum systems by on-chip fabrication, providing a universal platform to interface a  variety of solid-state qubit systems. For example, phonons have been used to mediate quantum gates between trapped ions\cite{Leibfried-IonReview-RevModPhys-2003}, coherently connect superconducting qubits\cite{Bienfait-PhononSCQubit-Science-2019} and to manipulate quantum-dot single-photon sources\cite{Yeo--QD-phonon-coupling--NNano-2014,Munsch2017}. The last decade has seen significant experimental effort towards coupling between spin qubits and mechanical motion\cite{Wang2020APL}.

Spins can be coupled to mechanics via induced strain in their host crystal or a oscillating magnetic moment \cite{Ghobadi-NVoptomechanicalInterfcace-PRA-2019}, and can be modeled using the canonical system shown in Fig.\,\ref{fig:interface}\,(b).  In recent experiments, coupling between piezoelectronically actuated mechanical motion and electronic spins has been realized for color centers in bulk diamond and SiC\cite{MacQuarrie-PRL-2013-StrainNVControl, Golter-PRX-2016-NVcoupledSAW, Maity-2020-NatComm-CoherentContolSiVacoustic, Whiteley-SiCspin-phononCoupling-NatPhys-2019}, in hybrid nanowire\cite{Arcizet-NVNanowire-2011-NatPhys, Pigeau-NVcouledToNanowireSpinProtection-NatCom-2015} and cantilever mechanical resonators (see Fig.\,\ref{fig:interface}\,(e))\cite{ref:teissier2014scn,Mesala-2016-PRAppl-NVmechanics, Ovartchaiyapong2014, Barfuss-StrongMechanicalDrivinfNV-NatPhys-2015}. However, combining these spin-mechanical devices with a coherent optical interface remains challenging owing to the weak interaction between photons and mechanical motion.

Cavity optomechanical devices provide a platform to enhance the spin-phonon-photon interaction by integrating mechanical resonators within optical cavities (See Sec.\,\ref{Sec:cavity_om}). In a cavity optomechanical system, resonant optical recirculation extends the photon-phonon interaction time, and a parametric enhancement of the optomechanical coupling rate, $g_{\textrm{om}}=\sqrt{\overline{n}_{\text{cav}}}g_0$, proportional to the intracavity photon number $\overline{n}_{\text{cav}}$ can be exploited.  A cavity optomechanical device can be designed to support optical modes in the telecommunication wavelength range, while simultaneously supporting mechanical modes resonant with the qubit spin transitions.  To maximize the spin-phonon and photon-phonon coupling rates, the devices can be engineered to minimize mechanical and optical mode volumes, respectively\cite{Mesala-2016-PRAppl-NVmechanics}. Nanoscale cavity optomechanical devices such as optomechanical crystals \cite{Burek2016} and microdisks \cite{Mitchell-DiamondOptomechanicalResonator-Optica-2016} typically support mechanical modes in the gigahertz frequency regime, which can be cooled to their quantum ground state \cite{Chan-LaserCoolingOptomecahnicsGroundState-Nature-2011, ref:teufel2011scm, Cohen-PhononCounting-Nature-2015} in a dilution fridge.

The operating principle of a spin-optomechanical interface based on spin-strain coupling is twofold, see Fig.\,\ref{fig:interface}\,(c):  radiation pressure from photons in an optical mode coherently excites the vibrations of a mechanical mode. This vibrational motion creates a microscopic stress field oscillating at the mechanical resonance frequency, which can interact with embedded spin qubits. The optomechanical interaction can be tuned for reversible photon-phonon conversion, and can operate at any wavelength resonant with a low-loss mode of the optical cavity. Recently, a proof-of-principle realization of a room-temperature cavity optomechanical spin-photon interface was demonstrated using a diamond microdisk resonator\cite{shandilya2021optomechanical}. In this work, a photonic coherent state in the 1,550\,nm telecommunication wavelength band was used to manipulate the electronic spin of a small ensemble of NV centers. Crucially, the resulting spin-photon interface does not rely on the intrinsic optical transitions of the color centers, thereby mitigating the aforementioned problems with spectral stability\cite{shandilya2021optomechanical}. Moreover, this approach is completely generic, and can be applied to color centers in other host materials\cite{Mitchell2014,shandilya2019hexagonal,das2021demonstration,Castelletto2020,Yan2021}, alongside providing a method to control optically inactive qubits\cite{Soykal-QubitsSiliconPhonons-PRL-2011,Degen2020}.

\subsection{Universal Optomechanical Spin Interface: Challenges and Solutions}

As previously discussed, quantum network applications require operation in a regime where a single photon coherently and reversibly couples to a spin qubit. In optomechanical devices, coherent spin-phonon interaction can be reached provided both the optomechanical- and the spin-phonon cooperativity $C_\text{om}$ and $C_\text{sm}$, respectively, exceed the thermal phonon number $n_{\textrm{th}}$. Cooling of the mechanical resonator to near the mechanical ground state ensures $n_{\textrm{th}}\ll1$ -- the following discussion will therefore ignore $n_{\textrm{th}}$. Assuming both the photon-phonon and the spin-phonon couplings are always `on', the transduction efficiency $\eta_\text{sp}$ of such a two-interface system is given by~\cite{Lauk-QuantScienceTechnology-2020-PerspectiveQuantumTransduction}:
\begin{equation}
    \eta_\text{sp} = \frac{4 C_\text{sm} C_\text{om} } {(1+ C_\text{sm} + C_\text{om})^2}\,.\label{eq:eta}
\end{equation}
A near-unity transduction efficiency is achievable in the limit $C_\text{sm} = C_\text{om}\gg1$\cite{Lauk-QuantScienceTechnology-2020-PerspectiveQuantumTransduction}. However, recent proposals lift the above restriction by employing temporal control of the coupling rates\cite{Mirhosseini2020,Neuman2021}.

The condition $C_\text{om} > 1$ has routinely been demonstrated in diamond optomechanical devices (see Sec.\,\ref{Sec:cavity_om}). However, an outstanding technical challenge is to perform coherent photon-phonon conversion in a device cryogenically cooled to near its mechanical quantum ground state, without heating due to optical absorption. Ground-state cooling, recently demonstrated in silicon optomechanical quantum memories\cite{Cohen-PhononCounting-Nature-2015, Wallucks-QuantumMemoryTelecomBand-NatPhys-2020}, will be aided by diamonds low nonlinear absorption and excellent thermal properties\cite{Mildren2013ch1}. 

To date, realizing $C_\text{sm} > 1$ has been hindered by the intrinsically weak spin-phonon coupling for the NV center ground state\cite{shandilya2021optomechanical}. However, in principle, reaching the coherent transduction regime is possible using already demonstrated diamond cavity optomechanical devices coupled to spin states with higher stress-sensitivity. For example, using SiV centers in microdisk resonators allows for $C_\text{sm}\sim1$, owing to their $10^5$-fold enhanced sensitivity to strain compared to the NV center ground state\cite{Wang2020APL}. Embedding SiV centers in optomechanical crystals will further increase $g_{\textrm{om}}$ on the account of smaller mechanical mode volume and a lower damping rate. Upon doing so, the regime $C_\text{sm}>100$ can be reached using already demonstrated spin qubits and devices\cite{Burek-DiamondOptomechanicalCrystal-Optica-2016,cady2019diamond, MacCabe-UltraLongPhononicLifetime-Science-2020,shandilya2021optomechanical}. Incorporating phononic shields can drastically reduce $\gamma_{\textrm{m}}$, potentially enabling $C_\text{sm}>10^4$, paving the way for deterministic quantum state transfer between single telecom photons and single spins\cite{shandilya2021optomechanical} and the use of phonons as on-chip quantum information carriers\cite{Lemonde2018,Zivari2022arxiv}. Working with SiV centers necessitates operation at mK temperatures. However, the NV center ground-state spins coupled via a phonon-assisted optical Raman process can reach similar high cooperativities at 8\,K\cite{Golter-PRX-2016-NVcoupledSAW}, albeit stabilizing the optical transition of NV centers in nanostructures presents a formidable challenge.

\section{Outlook: Enabling Quantum Networks with Diamond Integrated Photonics}

\begin{figure*}[tb]
\centering
	\includegraphics[width=\textwidth]{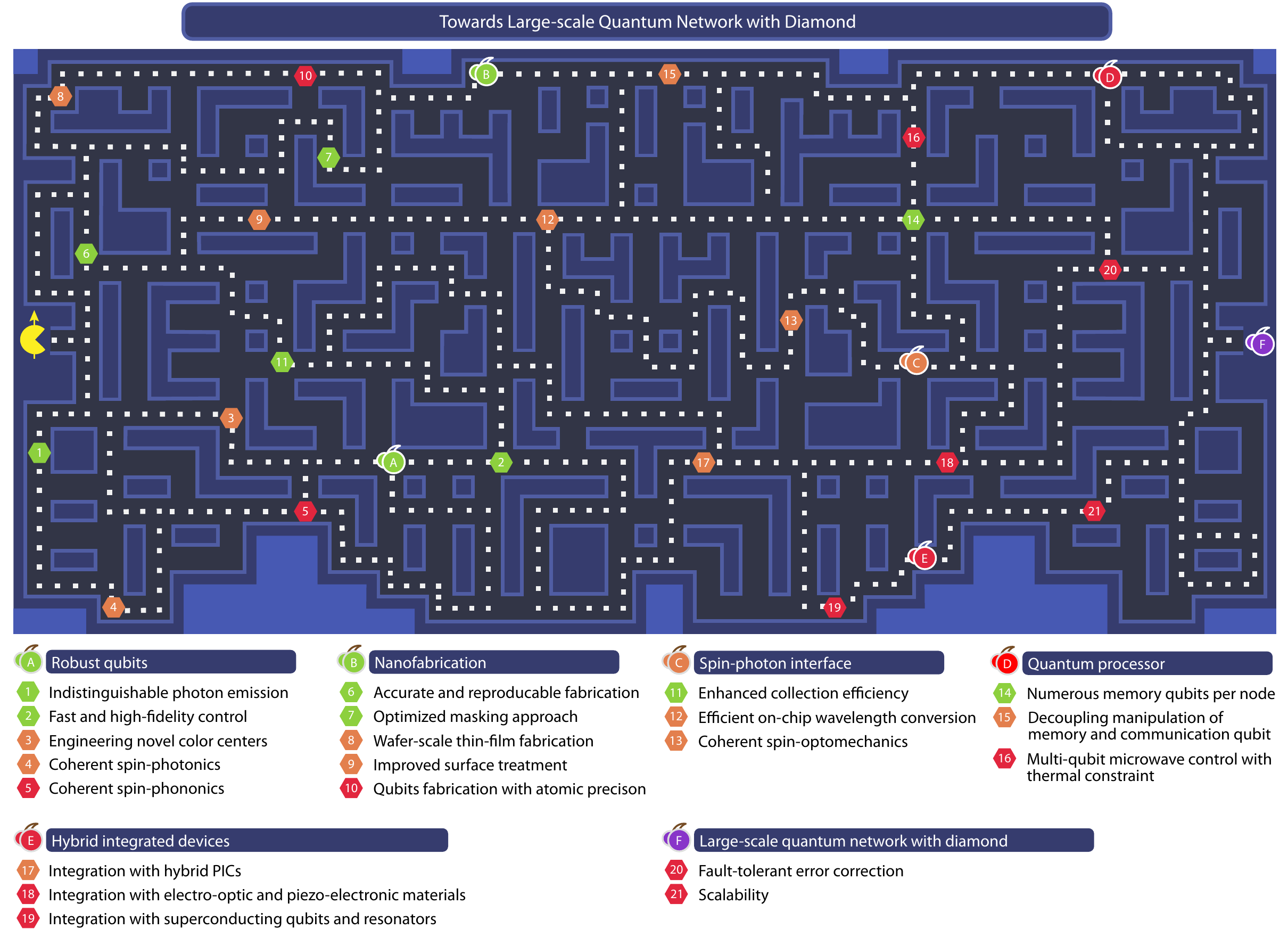}
		\caption{\textbf{Roadmap to the realization of  a large-scale quantum network with diamond:} The objectives have been broadly categorized into six categories, each section discussed briefly in the text. Subjective difficulty of each task\,/\,category is represented by the color of the coins/cherries, with green being the easiest and red being the hardest.}
		\label{fig:futureGoals}
\end{figure*}

Developments in diamond photonics have advanced tremendously over the years. However, to realize a fully functional quantum network, and for related quantum technologies to reach their maximum potential, there are numerous obstacles to overcome. A roadmap highlighting imminent challenges and future objectives in relation to  quantum hardware elements is presented in Fig.\,\ref{fig:futureGoals}. The following discussion is centered around addressing these challenges within the context of integrated diamond photonics.

\subsection{Robust Qubits}

For quantum networking protocols using NV centers \cite{Hensen-Nature-2015-LoopHoleFreeBell1p3km,Kalb-Science-2017-EntanglementDistillationNV,Humphreys-Nature-2018-RemoteEntanglement,pompili2021realization}, scalability is limited by the low flux of coherent photons. While, in principle, resonant coupling to a cavity enhances the photon flux\cite{Riedel2017}, NV centers embedded in nanophotonic cavities typically suffer from inhomogeneous linewidth broadening\cite{Faraon2011}, compromising photon indistinguishability and limiting the achievable entanglement rates, thus necessitating tuning and stabilizing of the ZPL transition\cite{Schmidgall2018,Acosta2012}. On the contrary, environment-insensitive color centers, such as SiV, perform excellently in nanophotonic resonators\cite{bhaskar2020experimental}. However, the SiV center experience heating-induced spin decoherence manifesting in slow and low-fidelity coherent microwave control\cite{Nguyen2019,Sukachev2017}. These limitations motivate the investigation of novel color centers.

As discussed briefly in Section\,\ref{Sec:CC_configuration}, careful control of defect concentration and boron doping, stabilizes the neutral SiV center, SiV$^0$\cite{Rose2018}. With a $S=1$ ground-state, SiV$^0$ combines the excellent spin properties of the NV center  ($T_2\sim900\,\textrm{ms}$ at 4\,K\cite{Rose2018}) with the favourable optical properties of SiV$^-$\cite{Zhang2020PRL,Rose2018}. Research on SiV$^0$ is still in the early stages, largely limited by the Fermi level pinning required to stabilize the charge state\cite{Rose2018}. 

\begin{table}[!b]
\renewcommand{\arraystretch}{2}
\caption{Optical Properties of Group-IV Color Centers in Diamond}
\centering
	\begin{tabular}{|c c c c c c|}
	    \hline
       \makecell{Color \\ center} & \makecell{ $\lambda_\text{zpl}$ \\ (nm)}  & \makecell{Radiative \\ lifetime (ns)} & \makecell{QE \\ ($\%$)} & \makecell{DW \\ factor} & \makecell{$\Delta_{\textrm{GS}}$ \\ (GHz)} \\
        \hline
        \hline
        SiV$^-$  & 738$^1$ & $\simeq 1.7-1.8$$^2$ & $1-10$$^3$ & 0.7$^4$ & 48$^5$ \\
        SiV$^0$  & 946$^6$ & $\simeq 1.8$$^7$ & \makecell{Not\\known}& 0.9$^8$ & N\,/A \\
        GeV$^-$ & 602$^9$ & $\simeq 1.4 - 6$$^{10}$ & 12-25$^{11}$ & 0.6$^{12}$ & $\sim160$$^{13}$\\
        SnV$^-$ &620$^{14}$  & $\simeq 4.5-7$$^{15}$ & $\sim$ 80$^{14}$ & 0.57$^{16}$ & 850$^{14}$\\
        PbV$^-$ & \makecell{520$^{17}$\\/\,550$^{18}$} & $\simeq$ $3-3.5$$^{17, 18}$ & \makecell{Not\\known} & \makecell{Not\\known} & \makecell{5700$^{17}$\\/\,4000$^{19}$}\\
        \hline
    \end{tabular}
\label{table_Group_IVsc}\\
$^1$\cite{Hepp2014}, $^2$\cite{Rogers2014,Sipahigil-Science-2016-SiVplatform}, $^3$\cite{Sipahigil-Science-2016-SiVplatform,Neu2012a}, $^4$\cite{Dietrich2014}, $^5$\cite{jahnke2015}, $^6$\cite{Rose2018}, $^7$\cite{Rose2018}, $^8$\cite{Rose2018}, $^9$\cite{Bhaskar2017a}, $^{10}$\cite{Iwasaki2015,Jensen2020}, $^{11}$\cite{Jensen2020,Nguyen2019GeV}, $^{12}$\cite{Bhaskar2017a}, $^{13}$\cite{Bhaskar2017a,Siyushev2017}, $^{14}$\cite{Iwasaki2017}, $^{15}$\cite{Trusheim-PRL-2020-TransformLimitedSnV,Gorlitz2020}, $^{16}$\cite{Gorlitz2020}, $^{17}$\cite{Trusheim-PRB-2019-PbV}, $^{18}$\cite{Tchernij2018,Froch2020}, $^{19}$\cite{Wang2021ACSPhotonics}.
\label{table_Group_IV}
\end{table}

For the negatively charged group-IV color centers, the ground-state splitting $\Delta_{\textrm{GS}}$ is found to increase with atomic number, owing to stronger spin-orbit interactions\cite{Thiering2018}. As a consequence, phonon-assisted population transfer between the orbital ground-states is suppressed, leading to prolonged spin coherence times at elevated temperatures\cite{Trusheim-PRL-2020-TransformLimitedSnV,Debroux2021,Bradac2019}. Table\,\ref{table_Group_IV} summarizes and compares the key physical properties of group-IV color centers. Note that research on the PbV center is still in the early stages and key parameters remain to be determined with consistency\cite{Wang2021ACSPhotonics}. 

Additionally, the reversible and coherent spin-photon coupling could, in principle, enable on-chip entanglement routing, thus potentially improving the communication rate\cite{lee2022quantum}. Alternatively, coupling of spins to propagating phonons has been proposed as a means to implement scalable, high-fidelity and hybrid on-chip quantum networks\cite{Lemonde2018,Zivari2022arxiv}.

\subsection{Nanofabrication}
The performance of integrated quantum photonic systems relies fundamentally on the engineering and nanofabrication of their key components.  Significant progress has been made in the nanostructuring of single-crystal diamond since the demonstration of nanowire antennas\cite{Babinec2010} and microring resonators\cite{Barclay2011, Faraon2011} over a decade ago to the recent realization of complex optomechanical devices integrated as spin-photon interfaces and waveguides in large-scale platforms\cite{Wan2020}. The lack of wafer-scale single-crystal diamond, though, limits the scalability of integrated photonic architectures and is a significant impediment to the realization of scalable fabrication analogous to what is achieved in the semiconductor industry. 

A key challenge in diamond nanostructuring is the mitigation of surface roughness that arises from the transfer of lithographically defined mask edge defects and mask erosion during plasma etching, as well as imperfections inherent to the diamond surface which have evolved from inhomogeneities on the seed substrate during CVD growth\cite{arnault2022}. This roughness can lead to degradation of both optical and acoustic (both bulk- and phononic-mode) quality factors, affecting the coupling between the two systems and hindering the transmission of information over long distances required for quantum communications applications. The fabrication of devices with dimensions required for smaller optical mode volumes is also affected by the prevalence of defects.  Efforts for surface roughness abatement prior to nanofabrication processes have included fine mechanical, chemical, and plasma-based smoothing\cite{Sangtawesin2019}, though the mechanisms for the anisotropic wear along different crystal facets during polishing are not yet well understood and therefore results can vary significantly based on methods\cite{hicks2019}.  There is also a necessity for the development of more robust, higher resolution electron beam resists with improved adhesion, as well as the optimization of hard masks and diamond plasma etching conditions.  

The creation of color centers with long coherence times at the surface or subsurface (within a few tens of nanometers) of diamond devices is an ongoing endeavor. Lattice defects induced by ion implantation as well as those inherent at the diamond surface create charge potentials that couple to the dipole moment of the spin\cite{ref:manson2005pin}, leading to dephasing between the two charge states and resulting in optical linewidth broadening compared to what is exhibited in the bulk\cite{Chakravarthi2021,Kasperczyk2020}. Several approaches have been made to understand and lessen the impacts of ion implantation, such as exposure through lithographically defined masks to minimize scattering\cite{Orwa2012} and post dosage annealing, though these methods have not been perfected.  Surface and subsurface damage removal has been accomplished through oxygen termination of the diamond surface post-annealing, improving optical linewidths and spin coherence by an order of magnitude\cite{Sangtawesin2019}.   

\subsection{Spin-Photon Interface}
The realization of a quantum network hinges on the interconnection of remote network nodes\cite{Kimble2008}. Cavity QED devices offer an ideal platform to interface stationary spin qubits with photons on account of the enhanced light-matter interactions\cite{ReisererRempe-RMP-2015-CavityQED}. These devices need to combine a large photon extraction efficiency\cite{bhaskar2020experimental} with the capability of tuning the emission frequency via on-chip application of electrical\cite{Bernien2013} or mechanical\cite{Sohn-2018-NatComm-SiVstrainControl,MacHielse2019} fields and microwaves for spin control. In addition, efficient on-chip frequency down-conversion from visible to telecommunication band is required to mitigate losses in the photonic links. Heralded entanglement schemes rely on quantum interference of indistinguishable photons. Quantum frequency down-conversion provides a method for interfering photons from spectrally distinguishable emitters, by detuning the pump laser to compensate for the spectral difference\cite{Stolk2022}. Alternatively, coherent photon-phonon and spin-phonon coupling in an optomechanical cavity provide a way to realize a spin-photon interface operating directly at telecom wavelengths\cite{shandilya2021optomechanical,Ghobadi-NVoptomechanicalInterfcace-PRA-2019,Raniwala2022}. 

\subsection{Quantum Processor}
A universal quantum processor consists of numerous qubits with long coherence times while supporting conditional and unconditional quantum gates\cite{Taminiau2014}. The realization of such a multi-qubit processor requires quantum gates to individually address the memory qubits, without introducing cross-talk affecting the coherence of the remaining spin-register\cite{Bradly-PRX-2019-NV10qubit1minuteCoherence}. Using dynamical-decoupling sequences\cite{Raiserer2016,Kalb2018,Abobeih-NatComm-2018-1sCoherenceTimeNV,Whaites2022} and selective coupling to $^{13}$C nuclear spins, a ten-qubit spin register based on the NV center has successfully been demonstrated\cite{Bradly-PRX-2019-NV10qubit1minuteCoherence}. While coupling to $^{13}$C nuclear spins has been demonstrated for the SiV center\cite{Nguyen2019,bhaskar2020experimental}, extending these results to multi-qubit registers remains challenging, owing to the spin-half nature of both the SiV and $^\text{13}$C spin systems that requires extended decoupling times\cite{Ruf2021JAP,Nguyen2019}. Realizing a multi-qubit register based on SiV or other group-IV color centers requires experimental effort to mitigate the aforementioned problem with heating-induced spin decoherence as a result of the applied microwave pulses.

\subsection{Hybrid Integrated Devices}

Hybrid integrated devices combine the strengths of disparate quantum systems to build complex quantum architectures. While individual monolithic diamond spin-photon interfaces have been realized with excellent performance\cite{Burek2017}, wafer-scale device fabrication remains elusive\cite{Wan2020}. Interfacing diamond with well-established photonic materials, such as AlN and GaP, provides a path towards incorporating diamond in photonic integrated circuits (PIC) for on-chip photon routing\cite{Wan2020}. Furthermore, using materials with strong $\chi^2$-nonlinearities enables on-chip frequency conversion\cite{Lake2016,Huang2021}. In addition, frequency shifting of single photons using electro-optical modulators paves the way for the entanglement of spectrally distinguishable quantum emitters\cite{Levonian2022}. 

Finally, piezo-electric materials have been proposed as a phononic interface between superconducting qubits and quantum memories based on color centers in diamond\cite{Neuman2021}. In this transduction scheme, a microwave photon is converted to a phonon via the piezoelectric effect, which further interacts with the quantum memory via spin-phonon coupling\cite{Kurokawa2022}. Quantum interference of the emitted photons thus enables entanglement of remote superconducting circuits. An advantage of this quantum memory-based transduction scheme over direct microwave-to-optical-photon conversion is a potential reduction in heating of the cryostat on account of the lower laser power required, which might preserve the coherence of the superconducting resonators\cite{Kurokawa2022}.

\subsection{Error Correction and Scalability}
Quantum systems are inherently noisy: interactions with the environment lead to decoherence, inevitably manifested by the emergence of errors\cite{waldherr2014quantum}. Correcting errors is therefore a necessity. Quantum error correction protocols have been demonstrated on spin registers based on NV centers in diamond\cite{Taminiau2014,waldherr2014quantum,Cramer2016,Unden2016}. These protocols harness the weak coupling between the electron spin and the surrounding nuclear spins. Recently, this development was pushed one step further by the demonstration of fault-tolerant operations on a diamond quantum processor\cite{Abobeih2022}. While in the early stages, the successful implementation of fault-tolerant operations has the potential to bring diamond to the forefront of quantum information processing based on solid-state spins.

At the time of writing, the scalability of diamond as a platform for quantum information processing is limited in part by challenges in nanofabrication and large-scale growth of single-crystal diamond\cite{Kianinia2020NatPhot}. A tremendous experimental effort is being invested to advance the state-of-the-art in diamond fabrication\cite{Mitchell-2019-APLphotonics-DiamondMicrodisks,Challier2018}, and the aforementioned hybrid platforms provide a promising route to realize scalable photonic platforms\,\cite{Wan2020}. In parallel, new 
techniques\cite{Schreck2017} are being pursued to increase the quality\cite{Nelz2019} and size\cite{Yamada2014,Kim2021DiamondGrowth} of synthetic diamonds. However, despite this progress large-scale synthetic growth of single-crystal diamond wafers remains elusive.

\subsection*{Concluding Remarks}
Research in the field of diamond integrated quantum photonics is vibrant and fast-moving, owing to diamond's unique combination of physical properties and ability to host robust spin qubits. In this article, we have reviewed the current state-of-the-art in the field, with particular emphasis on advances and outstanding challenges in nanofabrication, cavity optomechanics, and the development of qubit-photon interfaces. To conclude, we provided a road map illuminating the path towards the realization of a universal quantum network.

\bibliographystyle{IEEEtran}

\end{document}